\documentclass[aps,pre,onecolumn,12pt,superscriptaddress,nofootinbib]{revtex4-2}
\usepackage{amsmath,amssymb,bm}
\usepackage{xcolor}
\usepackage{graphicx}
\usepackage{hyperref}

\newcommand{\Tr}{\mathrm{Tr}}
\newcommand{\E}{\mathbb{E}}
\newcommand{\RR}{\mathsf{R}}
\newcommand{\cG}{\mathcal{G}}
\newcommand{\cC}{\mathcal{C}}
\newcommand{\cD}{\mathcal{D}}

\begin{document}

\title{Synchronous and asynchronous Monte Carlo dynamics of mean-field spin glasses\\
with rotationally invariant couplings: what they share and where they differ}

\author{Yoshiyuki Kabashima}
\affiliation{Institute for Physics of Intelligence, The University of Tokyo, Tokyo 113-0033, Japan}
\affiliation{Trans-Scale Quantum Science Institute, The University of Tokyo, Tokyo 113-0033, Japan}

\date{\today}

\begin{abstract}
We compare two Markov chain Monte Carlo dynamics for Ising spin glasses with rotationally invariant couplings:
the synchronous Monte Carlo method (SyncMC), which updates all spins in parallel with the help of Gaussian auxiliary fields,
and asynchronous Metropolis--Hastings dynamics, which updates one spin at a time.
Using the Guionnet--Ma\"ida formula, we derive the dynamical mean-field theory of both dynamics in terms of the R-transform of the coupling spectrum.
The two effective single-spin processes share a retarded self-interaction and a Gaussian noise;
SyncMC has, in addition, an instantaneous self-coupling and a noise that depends only on the response function.
These additional terms act only on the fastest time scale, so that the equations for slow time scales, including the fluctuation--dissipation ratio and, for the Sherrington--Kirkpatrick model, the Parisi equation, are the same for the two dynamics.
Simulations of the random orthogonal and Sherrington--Kirkpatrick models confirm that both dynamics approach the same fluctuation--dissipation relations, differing only in how many updates they need to do so.
\end{abstract}

\maketitle

\section{Introduction}

Markov chain Monte Carlo (MCMC) sampling of spin systems is usually carried out with asynchronous updates:
one spin is chosen at a time and updated according to the Metropolis--Hastings (MH) or heat-bath rule.
Such dynamics satisfy detailed balance with respect to the Boltzmann distribution.
On modern parallel hardware, however, updating one spin at a time is inefficient.
The synchronous Monte Carlo method (SyncMC)~\cite{Kabashima2026PRL,Kabashima2026JSTAT} updates all spins in parallel
by introducing Gaussian auxiliary fields for pairwise interactions, while still sampling the Boltzmann distribution exactly.
This raises a natural question: which aspects of the dynamics are common to SyncMC and asynchronous MH, and which are specific to each?
Answering this question requires a description in which both dynamics can be compared term by term.

For mean-field models, dynamical mean-field theory (DMFT) provides such a description: in the limit of a large number of spins $N$, it reduces the dynamics of the whole system to that of a single spin in an effective field.
The dynamics of mean-field spin glasses has been studied extensively by DMFT, mostly for soft or spherical spins obeying Langevin equations~\cite{SompolinskyZippelius1982,CugliandoloKurchan1993,CugliandoloKurchan1994}.
For Ising spins, the DMFT of single-spin-flip dynamics of the Sherrington--Kirkpatrick (SK) model~\cite{SK1975} was formulated by Sommers~\cite{Sommers1987},
and Eissfeller and Opper~\cite{EissfellerOpper1992} solved the effective single-spin process of the deterministic parallel dynamics $S_i(t+1)=\mathrm{sgn}\,h_i(t)$ (zero temperature, no Boltzmann sampling) by Monte Carlo sampling of single-spin trajectories.
They noted that it was not clear whether a similar single-spin equation can be derived for random sequential updates; for the Metropolis--Hastings rule, Sec.~\ref{sec:single} and Appendix~\ref{app:cont} of the present paper show how random sequential trials become a continuous-time single-spin process as $N\to\infty$.
For SyncMC, the DMFT was derived for the SK model~\cite{Kabashima2026JSTAT} and for the random orthogonal model (ROM)~\cite{MPR1994,Kabashima2026PRL}.

In this paper we derive the DMFT of SyncMC and of asynchronous MH in a common form, for Ising spin glasses whose coupling matrix is drawn from a rotationally invariant ensemble.
This class includes the SK model and the ROM.
Our approach rests on the Guionnet--Ma\"ida formula~\cite{GuionnetMaida2005}, which expresses the average of a finite-rank spherical integral
through the R-transform of the eigenvalue distribution. Because in both dynamics the couplings enter the dynamical generating functional
only through a finite-rank term, the formula yields DMFT effective single-spin processes in closed form for an arbitrary spectrum.

The main results are as follows.
\begin{enumerate}
\item The effective processes of the two dynamics share a retarded self-interaction $\RR(\cG)-\kappa_1$ and a Gaussian noise with covariance
$\Sigma_\phi[\cG,\cC]$ [Eqs.~\eqref{eq:MH_eff} and \eqref{eq:Sync_eff}].
Here $\RR$ is the R-transform of the eigenvalue distribution of the coupling matrix, $\kappa_1$ is the mean of that distribution,
and $\cG$ and $\cC$ are the response and correlation functions (Sec.~\ref{sec:model}).
\item SyncMC has, in addition, an instantaneous self-coupling of strength $\kappa_1+c$, where $c$ is the constant added to the diagonal of the coupling matrix in SyncMC, and a Gaussian noise whose covariance
$\beta^{-1}\cD[\cG]$ depends only on the response function [Eq.~\eqref{eq:D}].
\item These SyncMC-specific terms affect only the fastest time scale. Under time-scale separation, the slow-time-scale equations of the two dynamics coincide (Sec.~\ref{sec:slow}).
\item Simulations of both dynamics, together with numerical solutions of the effective processes, support this picture (Sec.~\ref{sec:results}).
In the paramagnetic phase the effective processes reproduce the finite-$N$ data of both dynamics.
Below the transition, the fluctuation--dissipation relations of both dynamics approach the same predictions: the one-step-aging form with $q=0.981$ and $x=0.679$ for the ROM at $T=0.1$, and the Parisi solution for the SK model at $T=0.5$.
For the same coupling matrices, the two dynamics give the same integrated response at fixed correlation when their waiting times are related by the time needed to reach the same energy;
one MCS of asynchronous MH then corresponds to about $10^2$--$10^3$ SyncMC steps for the ROM at $T=0.1$ and to $15$--$18$ steps for the SK model at $T=0.5$.
\end{enumerate}
The update rule thus sets how fast a system moves through its configuration space, but not which states it visits on slow time scales.
In practice, slow relaxation and aging of these models can be studied with either dynamics.
SyncMC needs more updates to reach a given state but performs each update for all spins in parallel; on a GPU this gives a shorter wall-clock time in our implementation, by a factor that depends on the model and the regime (Sec.~\ref{sec:numerics}).

The paper is organized as follows. Section~\ref{sec:model} defines the model and the two dynamics.
Section~\ref{sec:GM} explains how the Guionnet--Ma\"ida formula is used.
Sections~\ref{sec:MH} and \ref{sec:Sync} derive the effective processes of asynchronous MH and SyncMC.
Section~\ref{sec:compare} compares them. Section~\ref{sec:slow} analyzes slow time scales.
Section~\ref{sec:numerics} describes numerical methods, and Sec.~\ref{sec:results} presents numerical results.
Section~\ref{sec:discussion} is devoted to discussion.

\section{Model and dynamics}
\label{sec:model}

\subsection{Model}

We consider $N$ Ising spins $\bm\sigma=(\sigma_1,\dots,\sigma_N)^\top$, $\sigma_i\in\{+1,-1\}$, with energy
\begin{equation}
H(\bm\sigma)=-\frac12\sum_{i\neq j}J_{ij}\sigma_i\sigma_j .
\label{eq:H}
\end{equation}
The symmetric coupling matrix is $J=O^\top DO$, where $O$ is an $N\times N$ orthogonal matrix drawn uniformly (Haar measure)
and $D=\mathrm{diag}(\lambda_1,\dots,\lambda_N)$ is a fixed diagonal matrix whose eigenvalue distribution converges to a density $\rho(\lambda)$ as $N\to\infty$.
The R-transform of $\rho$ is denoted by $\RR(z)=\sum_{n\ge0}\kappa_{n+1}z^n$, where the coefficients $\kappa_n$ are the free cumulants of $\rho$.
It is defined by $\RR(g(z))+1/g(z)=z$ with $g(z)=\int\rho(\lambda)\,d\lambda/(z-\lambda)$.
We write $\mathcal R(x)=\int_0^x\RR(v)\,dv$. The inverse temperature is $\beta$.
For any constant $c$, the spectrum of $J+cI$ has R-transform $\RR(z)+c$; that is, only $\kappa_1$ is shifted by $c$.

Two examples are used throughout.
For the SK model, the couplings $J_{ij}=J_{ji}$ ($i<j$) are drawn independently from $\mathcal N(0,1/N)$, the normal distribution with mean $0$ and variance $1/N$, as in Ref.~\cite{Kabashima2026JSTAT}.
The coupling matrix then belongs to the Gaussian orthogonal ensemble, which is rotationally invariant (the diagonal elements do not enter $H$),
and $\RR(z)=z$, that is, $\kappa_2=1$ and $\kappa_n=0$ for $n\neq2$.
For the ROM, the eigenvalues of $D$ are $\pm1$ with equal weight and
\begin{equation}
\RR(z)=\frac{\sqrt{1+4z^2}-1}{2z}.\label{eq:R_ROM}
\end{equation}

\subsection{Asynchronous Metropolis--Hastings dynamics}

The dynamics proceeds by trials. In each trial, one spin $i$ is chosen uniformly at random and flipped with probability $\min(1,e^{-2\beta\sigma_ih_i})$, where
$h_i=\sum_{j\neq i}J_{ij}\sigma_j$ is the local field. A single trial thus updates at most one spin.
We let $n=0,1,2,\dots$ count the trials and define the time as $t=n/N$, so that one unit of time, one Monte Carlo step (MCS), consists of $N$ trials.
In one MCS each spin is chosen once on average, but the number of times a given spin is chosen fluctuates: it is Poisson distributed with mean one as $N\to\infty$.
In this limit, $t$ becomes a continuous variable and each spin becomes a continuous-time jump process with flip rate $W(\sigma,h)=\min(1,e^{-2\beta\sigma h})$.

\subsection{Synchronous Monte Carlo}

Let $A=J+cI$, where the diagonal shift $c$ is chosen so that $A$ is positive semidefinite (for the ROM, $c=1$).
One SyncMC step consists of two parallel updates:
\begin{align}
\bm x(t)&=A\bm\sigma(t)+\beta^{-1/2}A^{1/2}\bm\zeta(t),\qquad \bm\zeta(t)\sim\mathcal N(0,I_N),\label{eq:sync_x}\\
P\big(\sigma_i(t+1)=\pm1\,\big|\,\bm x(t)\big)&=\frac{e^{\pm\beta x_i(t)}}{2\cosh\beta x_i(t)}\quad\text{independently for all }i.\label{eq:sync_s}
\end{align}
Here $\mathcal N(0,I_N)$ denotes the standard normal distribution in $\mathbb R^N$.
The auxiliary field $\bm x$ has the units of energy and plays the role of the local field, in the same way as $h_i$ in asynchronous MH.
This is a blocked Gibbs sampler for the joint distribution
$p(\bm\sigma,\bm x)\propto\exp\!\big(-\beta\,\bm x^\top A^{-1}\bm x/2+\beta\,\bm\sigma^\top\bm x\big)$,
whose marginal for $\bm\sigma$ is the Boltzmann distribution of Eq.~\eqref{eq:H} because $\sigma_i^2=1$.
Hence the chain of $\bm\sigma$ is reversible with respect to the Boltzmann distribution.
Here $t=0,1,2,\dots$ counts SyncMC steps; it is an integer, and each step updates all $N$ spins simultaneously.
This is the SyncMC of Refs.~\cite{Kabashima2026PRL,Kabashima2026JSTAT}, written with the auxiliary field in units of energy.
In Ref.~\cite{Kabashima2026PRL} the field $\bm x$ is denoted $\bm u$ and the matrix $A$ is denoted $K$; in Ref.~\cite{Kabashima2026JSTAT} the field is denoted $\bm h$.
In both references the response function, denoted $G$ here, is denoted $R$; we reserve $\RR$ for the R-transform.

\subsection{Time units}
\label{sec:time}

The time $t$ has different meanings in the two dynamics.
In asynchronous MH, $t$ is a continuous variable measured in MCS; one unit of $t$ consists of $N$ trials, each of which updates at most one spin.
In SyncMC, $t$ is an integer counting steps; one unit of $t$ is a single step in which all $N$ spins are updated in parallel.
In both cases one unit of $t$ contains $N$ single-spin update opportunities, so we compare the two dynamics with one MCS of asynchronous MH corresponding to one SyncMC step.
The difference lies in how these opportunities are distributed: in SyncMC every spin is updated exactly once per unit time, all at the same instant,
whereas in asynchronous MH each spin is chosen at random instants whose number per unit time is Poisson distributed with mean one.

This difference carries over to the DMFT. For asynchronous MH, $C(t,s)$ and $G(t,s)$ are functions of two continuous times;
$G(t,s)$ is a response per unit time, and products of the corresponding operators are time integrals.
For SyncMC, $C(t,s)$ and $G(t,s)$ are defined on integer times; $G(t,s)$ is the response to a field applied during one step, and operator products are sums over intermediate steps.
With this convention, the expressions written in terms of operators below apply to both dynamics.
Physical quantities that depend on how time is measured, such as relaxation times, are compared in units of MCS and SyncMC steps, respectively.

In both dynamics we assume that the initial configuration is drawn independently of $J$.
An external field $\theta_i(t)$ conjugate to $\sigma_i$ (entering the energy as $-\sum_i\theta_i\sigma_i$) is used to define the response function
$G(t,s)=\delta\langle\sigma(t)\rangle/\delta\theta(s)$. In SyncMC it enters as $x_i(s)\to x_i(s)+\theta_i(s)$.

\section{Generating functional}
\label{sec:GM}

The generating functional of each dynamics is a sum over spin trajectories, weighted by their probability, in which the local field is introduced as an integration variable and its definition is imposed by auxiliary (conjugate) variables.
It consists of two parts: a part containing the couplings, which we average over the rotation $O$ (Sec.~\ref{sec:rot}),
and a part that factorizes over sites once the order parameters are introduced (Sec.~\ref{sec:single}).
How each part is brought to a form suitable for $N\to\infty$ differs between the two dynamics, because their times have different meanings (Sec.~\ref{sec:time}).

\subsection{Average over the rotation}
\label{sec:rot}

For a real orthogonal Haar matrix $O$ and a complex matrix $B$ of fixed finite rank,
\begin{equation}
\lim_{N\to\infty}\frac1N\log\E_O\exp\!\big(N\,\Tr(B\,O^\top MO)\big)=\frac12\Tr\,\mathcal R_M(2B),
\label{eq:GM}
\end{equation}
where $\E_O$ is the average over $O$, $M$ is a diagonal matrix with limiting spectrum whose integrated R-transform is $\mathcal R_M$,
and $\mathcal R_M$ acts on the nonzero eigenvalues of $2B$~\cite{GuionnetMaida2005}.
The formula is proved for real arguments; we use it for complex $B$ by analytic continuation.

Equation~\eqref{eq:GM} requires that the rank of $B$ stay finite as $N\to\infty$.
This requirement is met differently in the two dynamics. Let $L$ denote the maximum time, in SyncMC steps or in MCS.
\begin{itemize}
\item In SyncMC, the dynamics consists of $L$ steps, and the path integral is written on the $T=L$ integer times. No further discretization is needed.
\item In asynchronous MH, the dynamics consists of $NL$ trials. A path integral written trial by trial would have $T=NL$ time points, which grows with $N$,
and Eq.~\eqref{eq:GM} could not be applied. We therefore introduce a coarse time grid with spacing $\Delta t$ (in MCS), independent of $N$, and impose the definition of the local field
only at the $T=L/\Delta t$ grid points, holding the field fixed within each interval. Each interval contains $N\Delta t$ trials,
and each spin is chosen a Poisson-distributed number of times with mean $\Delta t$ within it.
The error from holding the field fixed vanishes as $\Delta t\to0$.
We take $N\to\infty$ at fixed $\Delta t$ first, and $\Delta t\to0$ afterwards, which yields the continuous-time description.
\end{itemize}
In both cases $T$ does not depend on $N$. After the dynamics is written as a path integral with auxiliary variables on these $T$ time points,
the couplings appear only through a factor $\exp(N\Tr(BA'))$, where $A'$ is $J$ or $J+cI$ and
\begin{equation}
2B=\frac1N\,XKX^\top,\qquad X=[\bm v(1),\dots,\bm v(T),\hat{\bm v}(1),\dots,\hat{\bm v}(T)]\in\mathbb R^{N\times 2T}.
\end{equation}
Here $\bm v$ is the dynamical variable on which the couplings act, $\hat{\bm v}$ is its conjugate auxiliary variable, and $K$ is a fixed $2T\times2T$ matrix.
The rank of $B$ is at most $2T$, independent of $N$.
The nonzero eigenvalues of $2B$ coincide with those of $KQ$, where $Q=X^\top X/N$ collects the order parameters, so that
\begin{equation}
\E_O\exp\big(N\Tr(BA')\big)\simeq\exp\Big(\frac N2\Tr\,\mathcal R_{A'}(KQ)\Big).
\end{equation}
At the saddle point, causality makes the diagonal blocks of $KQ$ strictly lower or strictly upper triangular in time.
Hence all eigenvalues of $KQ$ vanish, and the arguments of $\mathcal R_{A'}$ stay in the neighborhood of the origin where Eq.~\eqref{eq:GM} applies.

\subsection{Single-site part and the continuous-time limit}
\label{sec:single}

We denote by $u_i(t)$ the local field of site $i$ ($u=h$ for asynchronous MH and $u=x$ for SyncMC) and by $\hat u_i(t)$ its conjugate variable.
The order parameters $Q$ are introduced through conjugate variables $\hat Q$, which gives
\begin{equation}
Z=\int dQ\,d\hat Q\,\exp\Big\{N\Big[\frac12\Tr\,\mathcal R_{A'}(KQ)-\frac12\Tr\,\hat QQ+\log Z_1[\hat Q]\Big]\Big\},
\end{equation}
where the single-site generating functional is
\begin{equation}
Z_1[\hat Q]=\sum_{\sigma(0),\dots,\sigma(T)}\int\prod_{k}\frac{du_k\,d\hat u_k}{2\pi}\;w[\sigma\,|\,u]\;
\exp\Big(i\sum_k\hat u_k\big(u_k-\theta_k+\epsilon\,\kappa_1\sigma_k\big)+\frac12\,\bm y^\top\hat Q\,\bm y\Big),
\label{eq:Z1}
\end{equation}
with $\bm y=(\sigma_1,\dots,\sigma_T,\hat u_1,\dots,\hat u_T)^\top$, $\epsilon=1$ for asynchronous MH and $\epsilon=0$ for SyncMC.
The subscript $k$ labels the $T$ time points of Sec.~\ref{sec:rot}, and $\sigma_k$, $u_k$, $\hat u_k$, $\theta_k$ are the values at these points.
The single-site path weight $w[\sigma|u]$, the probability of the spin trajectory given the history of the local field, is where the two dynamics differ.

\paragraph{SyncMC.}
The path weight follows directly from Eq.~\eqref{eq:sync_s}:
\begin{equation}
w_{\rm S}[\sigma\,|\,x]=p_0\big(\sigma(0)\big)\prod_{t=0}^{L-1}\frac{e^{\beta\sigma(t+1)x(t)}}{2\cosh\beta x(t)},
\label{eq:wS}
\end{equation}
where $p_0$ is the initial distribution. This expression is exact for every $N$, the time points are the $T=L$ steps, and no limit in time is taken.
The transition probability from $\sigma(t)$ to $\sigma(t+1)$ does not depend on $\sigma(t)$ except through $x(t)$: each step resamples the spin.

\paragraph{Asynchronous MH.}
Consider the interval $[t_k,t_{k+1})$ of the coarse grid, $t_k=k\Delta t$, during which the field is held at $h_k$.
The interval contains $N\Delta t$ trials, and a given site is chosen in each of them with probability $1/N$.
The number of times the site is chosen is therefore binomially distributed and becomes Poisson distributed with mean $\Delta t$ as $N\to\infty$.
Each time the site is chosen, its spin is updated by the single-trial transition matrix $P_1(h_k)=I+\mathbb W(h_k)$, where $\mathbb W(h)$ is the $2\times2$ matrix with elements
\begin{equation}
\mathbb W_{-\sigma,\sigma}(h)=W(\sigma,h),\qquad \mathbb W_{\sigma,\sigma}(h)=-W(\sigma,h)
\end{equation}
(the first index is the new state). Averaging over the Poisson-distributed number $m$ of choices gives
\begin{equation}
\sum_{m\ge0}e^{-\Delta t}\frac{(\Delta t)^m}{m!}\,P_1(h_k)^m=e^{\Delta t\,\mathbb W(h_k)},
\end{equation}
so that the path weight on the coarse grid is
\begin{equation}
w_{\rm MH}[\sigma\,|\,h]=p_0\big(\sigma(0)\big)\prod_{k=0}^{T-1}\Big[e^{\Delta t\,\mathbb W(h_k)}\Big]_{\sigma_{k+1},\sigma_k}.
\label{eq:wMH}
\end{equation}
Two approximations enter: the field is held fixed within each interval, and the numbers of choices of different sites are treated as independent (their correlation is of order $1/N$).
The first vanishes as $\Delta t\to0$ and the second as $N\to\infty$.
Since $e^{\Delta t\,\mathbb W}=I+\Delta t\,\mathbb W+O(\Delta t^2)$, the product in Eq.~\eqref{eq:wMH}, divided by $(\Delta t)^n$ for a trajectory with $n$ flips, converges as $\Delta t\to0$ to the probability density of a continuous-time jump-process trajectory
that flips at times $\tau_1<\tau_2<\cdots<\tau_n$ (Appendix~\ref{app:cont}):
\begin{equation}
w_{\rm MH}[\sigma\,|\,h]\to p_0\big(\sigma(0)\big)\exp\Big(\sum_j\log W\big(\sigma(\tau_j^-),h(\tau_j)\big)-\int_0^L W\big(\sigma(t),h(t)\big)\,dt\Big),
\label{eq:wMHcont}
\end{equation}
where $\tau_j^-$ is the instant just before the $j$-th flip.
This is a density with respect to $d\tau_1\cdots d\tau_n$; correspondingly, the sum over trajectories on the grid becomes
\begin{equation*}
\sum_{\sigma(0)}\sum_{n\ge0}\int_{0<\tau_1<\cdots<\tau_n<L}d\tau_1\cdots d\tau_n .
\end{equation*}
In contrast to SyncMC, the transition matrix $e^{\Delta t\,\mathbb W}$ is close to the identity: the spin keeps its value unless it is flipped.

\paragraph{Order parameters in the continuous-time limit.}
For asynchronous MH, a field $\theta_l$ applied during the interval $[t_l,t_{l+1})$ changes $\langle\sigma_k\rangle$ by an amount proportional to $\Delta t$.
We therefore write the response on the grid as $G_{kl}=\Delta t\,G(t_k,t_l)$, where $G(t,s)$ is the response per unit time, while $C_{kl}=C(t_k,t_l)$ needs no rescaling.
A product of $n$ response matrices on the grid then becomes $\Delta t$ times an $n$-fold time integral, and sums over grid points become integrals.
In this way the operator expressions in Secs.~\ref{sec:MH} and \ref{sec:Sync} hold for both dynamics, with time integrals for asynchronous MH and sums over steps for SyncMC.

\section{DMFT of asynchronous MH}
\label{sec:MH}

Imposing $h_i=(J\bm\sigma)_i-J_{ii}\sigma_i$ at the grid points of Sec.~\ref{sec:rot} with auxiliary variables $\hat h_i(t)$ and using $J_{ii}\to\kappa_1$ gives
$\bm v=\bm\sigma$, $\hat{\bm v}=\hat{\bm h}$ and $K=-i\Lambda$ with $\Lambda=\begin{pmatrix}0&I_T\\I_T&0\end{pmatrix}$.
Evaluating the saddle point with the single-site weight $w_{\rm MH}$ of Eq.~\eqref{eq:wMH} (details in Appendix~\ref{app:MH}) and taking $\Delta t\to0$ as in Sec.~\ref{sec:single} yields the following effective process.
A single spin flips with rate $W(\sigma(t),h(t))$, where
\begin{equation}
h(t)=\int_0^t\Gamma_{\rm MH}(t,s)\,\sigma(s)\,ds+\phi(t)+\theta(t),\qquad \Gamma_{\rm MH}=\RR(\cG)-\kappa_1=\sum_{n\ge1}\kappa_{n+1}\cG^n,
\label{eq:MH_eff}
\end{equation}
and $\phi$ is a Gaussian process with zero mean and covariance
\begin{equation}
\Sigma_\phi[\cG,\cC]=\sum_{n\ge1}\kappa_{n+1}\sum_{k=0}^{n-1}\cG^k\,\cC\,(\cG^\top)^{n-1-k}.
\label{eq:Sigma_common}
\end{equation}
Here time $t$ is continuous and measured in MCS (Sec.~\ref{sec:time}); $\cG$ and $\cC$ are the integral operators with kernels $G(t,s)$ and $C(t,s)$, products are time integrals, and $\cG^0$ is the identity.
The functions $C(t,s)=\langle\sigma(t)\sigma(s)\rangle$ and $G(t,s)=\delta\langle\sigma(t)\rangle/\delta\theta(s)$ are determined self-consistently,
where $\langle\cdots\rangle$ denotes the average over $\phi$ and over the jump process.
For the SK model, Eqs.~\eqref{eq:MH_eff} and \eqref{eq:Sigma_common} give $\Gamma_{\rm MH}=\cG$ and $\Sigma_\phi=\cC$.
The effective field then has the same structure as in the effective process of Sompolinsky and Zippelius~\cite{SompolinskyZippelius1982}:
a retarded self-interaction given by the response function and a Gaussian noise whose covariance is the correlation function.
The dynamics of the spin itself is different, however. Sompolinsky and Zippelius treated soft spins obeying a Langevin equation,
whereas here an Ising spin flips with the MH rate; the resulting $C$ and $G$ are therefore different functions, and only the structure of the effective field is shared.

\section{DMFT of SyncMC}
\label{sec:Sync}

Time $t=0,1,2,\dots$ counts SyncMC steps (Sec.~\ref{sec:time}); operator products are sums over intermediate steps. We impose Eq.~\eqref{eq:sync_x} with auxiliary variables $\hat{\bm x}(t)$ and integrate out $\bm\zeta(t)$.
The coupling-dependent factor becomes
\begin{equation}
\exp\Big(-i\sum_t\hat{\bm x}(t)^\top A\bm\sigma(t)-\frac{1}{2\beta}\sum_t\hat{\bm x}(t)^\top A\hat{\bm x}(t)\Big)=\exp\big(N\Tr(BA)\big),
\end{equation}
which has the form of Sec.~\ref{sec:rot} with $\bm v=\bm\sigma$, $\hat{\bm v}=\hat{\bm x}$ and
\begin{equation}
K=\begin{pmatrix}0&-iI_T\\-iI_T&-\beta^{-1}I_T\end{pmatrix}.
\end{equation}
The new feature compared with MH is the lower-right block $-\beta^{-1}I_T$, which comes from the noise term in Eq.~\eqref{eq:sync_x}.
Apart from this block, $K$ coincides with that of asynchronous MH, because $\bm x$ and $\bm h$ have the same units.

Since $\theta$ enters as $x\to x+\theta$, the response function satisfies $\langle\sigma(t)\hat x(s)\rangle=iG(t,s)$ at the saddle point, where also $\langle\hat x\hat x\rangle=0$. Then
\begin{equation}
KQ=\begin{pmatrix}\cG^\top&0\\-i(\cC+\beta^{-1}\cG^\top)&\cG\end{pmatrix}.
\end{equation}
Evaluating the stationarity conditions in the same way as for MH, now with the single-site weight $w_{\rm S}$ of Eq.~\eqref{eq:wS} and without any limit in time (Appendix~\ref{app:Sync}), yields the following effective process.
A single spin is updated by the heat-bath rule, $P(\sigma(t+1)=\pm1)\propto e^{\pm\beta x(t)}$, with
\begin{equation}
x(t)=\underbrace{(\kappa_1+c)\,\sigma(t)}_{\text{SyncMC-specific}}+\sum_{s<t}\Gamma_{\rm MH}(t,s)\,\sigma(s)+\phi(t)+\underbrace{\eta(t)}_{\text{SyncMC-specific}}+\theta(t).
\label{eq:Sync_eff}
\end{equation}
Here $\phi$ has the same covariance $\Sigma_\phi[\cG,\cC]$ as in Eq.~\eqref{eq:Sigma_common}.
The process $\eta$ is a Gaussian process independent of $\phi$, with zero mean and covariance $\beta^{-1}\cD[\cG]$, where
\begin{equation}
\cD[\cG]=\sum_{n\ge0}\kappa^{A}_{n+1}\sum_{j=0}^{n}\cG^j(\cG^\top)^{n-j}
=(\kappa_1+c)I+\kappa_2(\cG+\cG^\top)+\kappa_3(\cG^2+\cG\cG^\top+\cG^{\top2})+\cdots,
\label{eq:D}
\end{equation}
with $\kappa^A_1=\kappa_1+c$ and $\kappa^A_n=\kappa_n$ for $n\ge2$.
Equivalently, as a double contour integral around the origin,
$\cD[\cG]=\oint\!\oint\frac{dz\,dw}{(2\pi i)^2}\frac{z\RR_A(z)-w\RR_A(w)}{z-w}(z-\cG)^{-1}(w-\cG^\top)^{-1}$, with $\RR_A(z)=\RR(z)+c$.

\paragraph{Checks.}
(i) For $J=0$, Eq.~\eqref{eq:Sync_eff} reduces to $x=c\sigma+\mathcal N(0,c/\beta)$, which is SyncMC for independent spins with self-coupling $c$.
(ii) For the SK model,
\begin{equation*}
x(t)=c\sigma(t)+\sum_{s<t}G(t,s)\sigma(s)+\phi(t)+\eta(t),
\end{equation*}
with $\langle\phi\phi\rangle=\cC$ and $\langle\eta\eta\rangle=\beta^{-1}[cI+\cG+\cG^\top]$.
This coincides with the effective process of SyncMC for the SK model in Ref.~\cite{Kabashima2026JSTAT} [Eqs.~(10)--(12) there, where $\phi+\eta$ is written as a single noise].
(iii) For a general spectrum, Eqs.~\eqref{eq:Sync_eff} and \eqref{eq:D} coincide with the effective process derived for the ROM in Ref.~\cite{Kabashima2026PRL}, Eqs.~(8) and (9) there, where the noise covariance $\Sigma_\phi+\beta^{-1}\cD$ is written as a single matrix.

\section{Common and specific parts}
\label{sec:compare}

Table~\ref{tab:compare} summarizes the comparison.

\begin{table}[t]
\caption{Effective single-spin processes of asynchronous MH and SyncMC. Here $\Gamma_{\rm MH}=\RR(\cG)-\kappa_1$, $\Sigma_\phi$ is given by Eq.~\eqref{eq:Sigma_common}, and $\cD$ by Eq.~\eqref{eq:D}.}
\label{tab:compare}
\begin{ruledtabular}
\begin{tabular}{lll}
 & Asynchronous MH & SyncMC\\
\hline
Time & continuous, in MCS ($N$ trials) & integer, in steps (all $N$ spins)\\
Update rule & jump with rate $\min(1,e^{-2\beta\sigma h})$ & heat bath with $e^{\beta\sigma x}$\\
Retarded self-interaction & $\Gamma_{\rm MH}$ & $\Gamma_{\rm MH}$ (common)\\
Noise depending on $C$ & $\Sigma_\phi[\cG,\cC]$ & $\Sigma_\phi[\cG,\cC]$ (common)\\
Instantaneous self-coupling & none & $(\kappa_1+c)\,\sigma(t)$\\
Noise depending only on $G$ & none & $\beta^{-1}\cD[\cG]$\\
\end{tabular}
\end{ruledtabular}
\end{table}

\paragraph{Origin of the specific terms.}
In asynchronous MH, the self-interaction $J_{ii}$ is removed from the local field, and $\kappa_1$ cancels against the constant term of $\RR(\cG)$.
In SyncMC, the auxiliary field contains the diagonal of $A$, which produces the instantaneous self-coupling,
and the auxiliary noise $\beta^{-1/2}A^{1/2}\bm\zeta$, which is correlated through $J$, produces the noise $\eta$.

\paragraph{Fluctuation--dissipation theorem.}
For a function $f(t,s)=f(\tau)$ that depends only on $\tau=t-s$, we write its Fourier transform as $\tilde f(\omega)=\int d\tau\,e^{i\omega\tau}f(\tau)$.
For time-translation-invariant $C$ and $G$, the common noise satisfies
\begin{equation}
\tilde\Sigma_\phi(\omega)=\tilde C(\omega)\,\frac{\RR\big(\tilde G(\omega)\big)-\RR\big(\tilde G(\omega)^*\big)}{\tilde G(\omega)-\tilde G(\omega)^*},
\label{eq:SigmaFourier}
\end{equation}
where $^*$ denotes complex conjugation (see Appendix~\ref{app:Fourier} for the derivation of this and the next equation).
The fluctuation--dissipation theorem (FDT) at inverse temperature $\beta_{\rm e}$ states that $G(\tau)=-\beta_{\rm e}\,C'(\tau)\,\Theta(\tau)$,
where $C'=dC/d\tau$ and $\Theta$ is the unit step function, or equivalently $\tilde C(\omega)=(2/\beta_{\rm e}\omega)\,\mathrm{Im}\,\tilde G(\omega)$.
We write $\beta_{\rm e}$ for a general inverse temperature, because the relation is used both with $\beta_{\rm e}=\beta$ (equilibrium and the fastest time scale) and with $\beta_{\rm e}=\beta x$ (slow time scales in the aging regime, Sec.~\ref{sec:slow} and Appendices~\ref{app:Fourier} and \ref{app:Langevin}).
If $C$ and $G$ satisfy it, Eq.~\eqref{eq:SigmaFourier} becomes
\begin{equation}
\tilde\Sigma_\phi(\omega)=\frac{2}{\beta_{\rm e}\omega}\,\mathrm{Im}\,\RR\big(\tilde G(\omega)\big);
\label{eq:FDTpreserve}
\end{equation}
that is, the induced noise and retarded self-interaction, whose Fourier transform is $\RR(\tilde G(\omega))-\kappa_1$, satisfy the FDT at the same temperature.
This property, which holds for any spectrum, is the key ingredient of Sec.~\ref{sec:slow}.

\paragraph{Discrete-time FDT of SyncMC.}
For SyncMC, time is discrete and the FDT takes a different form.
In equilibrium, the response function of SyncMC satisfies, exactly and for any $N$~\cite{Kabashima2026JSTAT},
\begin{equation}
G(t,s)=\beta\big[C(t,s+1)-C(t,s)\big]\qquad(t\ge s+1),
\label{eq:FDT_discrete}
\end{equation}
where $G(t,s)$ is the response of $\langle\sigma(t)\rangle$ to a field $\theta(s)$ entering as $x(s)\to x(s)+\theta(s)$, which first affects $\sigma(s+1)$.
With $\tau=t-s$ and the discrete Fourier transform $\tilde f(\omega)=\sum_\tau e^{i\omega\tau}f(\tau)$ ($-\pi<\omega\le\pi$), Eq.~\eqref{eq:FDT_discrete} reads
\begin{equation}
\beta\sin\frac\omega2\;\tilde C(\omega)=\mathrm{Im}\big[e^{-i\omega/2}\,\tilde G(\omega)\big],
\label{eq:FDT_discrete_F}
\end{equation}
which reduces to the continuous-time form for $|\omega|\ll1$, that is, on time scales much longer than one step.

The SyncMC-specific noise $\eta$ is what makes the effective process consistent with Eq.~\eqref{eq:FDT_discrete}.
Consider the field $x$ of Eq.~\eqref{eq:Sync_eff} as a bath acting on the spin. Its response to the spin is $K=(\kappa_1+c)I+\Gamma_{\rm MH}$, with Fourier transform $\tilde K(\omega)=\RR_A(\tilde G(\omega))$,
and the covariance of its fluctuations is $\mathcal N=\Sigma_\phi+\beta^{-1}\cD$.
If $C$ and $G$ satisfy Eq.~\eqref{eq:FDT_discrete}, then
\begin{equation}
K(t,s)=\beta\big[\mathcal N(t,s)-\mathcal N(t,s-1)\big]\quad(t\ge s),
\label{eq:FDT_bath_discrete}
\end{equation}
or equivalently $\beta\sin(\omega/2)\,\tilde{\mathcal N}(\omega)=\mathrm{Im}\big[e^{i\omega/2}\,\RR_A(\tilde G(\omega))\big]$,
for any spectrum (Appendix~\ref{app:Fourier}). This is the discrete-time FDT of the bath, with the time shifted by one step relative to Eq.~\eqref{eq:FDT_discrete} because the field at time $t$ acts on the spin at time $t+1$.
Without the term $\beta^{-1}\cD$, $\Sigma_\phi$ and $\Gamma_{\rm MH}$ alone do not satisfy a discrete-time relation of this kind.
For the SK model, $\RR_A(z)=c+z$ and Eq.~\eqref{eq:FDT_bath_discrete} can be checked directly: $\mathcal N(t,s)=C(t,s)+\beta^{-1}[c\,\delta_{ts}+G(t,s)+G(s,t)]$, which by Eq.~\eqref{eq:FDT_discrete} equals $C(t,s+1)$ for $t\ge s+1$.
On slow time scales, $\cD$ does not contribute (Sec.~\ref{sec:slow}), and $\Sigma_\phi$ and $\Gamma_{\rm MH}$ satisfy the continuous-time relation \eqref{eq:FDTpreserve}, with $\beta_{\rm e}=\beta x$ in the aging regime.

\section{Slow time scales}
\label{sec:slow}

We now assume time-scale separation: the correlation function decays from $1$ to a plateau value $q$ on a fast time scale of order one,
and from $q$ to smaller values on slow time scales that grow with the waiting time.

\subsection{Why the slow-time-scale equations coincide}

In the derivations of slow-time-scale equations, the dynamics enters only through two quantities.
\begin{enumerate}
\item The local equilibrium of the spin together with the fast part of the noise and retarded self-interaction, at a fixed slow field $h$:
$P(\sigma|h)\propto e^{\beta h\sigma}$, with local magnetization $m=\tanh\beta h$.
\item The integrated response of the fast part, $\chi_F=\int G_F=\beta(1-q)$, which follows from the FDT on fast time scales.
\end{enumerate}
Both follow from detailed balance with respect to the Boltzmann distribution, which holds for asynchronous MH and for SyncMC.

The SyncMC-specific terms do not alter these two quantities, for two reasons.
\begin{itemize}
\item \emph{The instantaneous self-coupling.} Together with the fast parts of the retarded self-interaction and of the noise, it forms a Gaussian bath that is in equilibrium at the inverse temperature $\beta$ for any $c$.
Integrating out this bath leaves the local equilibrium $P(\sigma|h)\propto e^{\beta h\sigma}$ unchanged; the argument is given in Sec.~\ref{sec:fastbath}.
\item \emph{The noise $\eta$ beyond its white part.} Its covariance contains only products of response functions.
Response functions, measured per unit of real time, vanish on slow time scales (they are inversely proportional to the waiting time),
whereas the correlation function stays of order one. Hence the covariance of $\eta$ vanishes at slow time separations, and $\eta$ belongs entirely to the fast time scale.
\end{itemize}
The slow parts of $\Gamma_{\rm MH}$ and $\Sigma_\phi$ are built from the same $\RR$ in both dynamics. Therefore the slow-time-scale equations coincide.

\subsection{Local equilibrium in SyncMC}
\label{sec:fastbath}

\paragraph{Fast part of the field.}
The field of Eq.~\eqref{eq:Sync_eff} contains, besides the self-coupling, fast components from the recent past, $s=t-1,t-2,\dots$, in the retarded term and in $\phi$; these must be treated together with the self-coupling.
We write $x(t)=h(t)+\xi(t)$, where $h$ is the slow part of the field and
\begin{equation*}
\xi(t)=(\kappa_1+c)\,\sigma(t)+\sum_{s<t}\Gamma_F(t,s)\,\sigma(s)+\phi_F(t)+\eta(t)
\end{equation*}
collects the fast parts.

\paragraph{Splitting $C$, $G$ and $\phi$.}
To define the fast parts, we write $C=C_S+C_F$, where $C_F$ decays from $1-q$ to zero on the fast time scale and $C_S$ is the rest, which equals $q$ throughout the fast time scale, and we write $G_F$ for the response on the fast time scale, whose sum over time is $\chi_F=\beta(1-q)$.
The rest of $G$, the slow response, is of the order of the inverse waiting time per step and is negligible over the fast time scale.
Since $\Sigma_\phi[\cG,\cC]$ of Eq.~\eqref{eq:Sigma_common} is linear in $\cC$, $\phi$ is the sum of two independent Gaussian processes with covariances $\Sigma_\phi[\cG,\cC_S]$ and $\Sigma_\phi[\cG,\cC_F]$.
The first is constant over the fast time scale and belongs to $h$.
The second is $\phi_F$; over the fast time scale its covariance is $\Sigma_{\phi,F}=\Sigma_\phi[\cG_F,\cC_F]$.
Likewise, $\Gamma_F=\RR(\cG_F)-\kappa_1$ is the retarded self-interaction over the fast time scale.

\paragraph{The fast bath is in equilibrium.}
The field $\xi$ acts on the spin as a Gaussian bath coupled through the energy $-\sigma\xi$, with response $K_F=(\kappa_1+c)I+\Gamma_F$ and fluctuation covariance $\mathcal N_F=\Sigma_{\phi,F}+\beta^{-1}\cD[\cG_F]$.
As usual under time-scale separation, we assume that $C_F$ and $G_F$ satisfy the discrete-time FDT \eqref{eq:FDT_discrete} at the inverse temperature $\beta$ of the dynamics.
The derivation of Eq.~\eqref{eq:FDT_bath_discrete} in Appendix~\ref{app:Fourier}, applied to $C_F$ and $G_F$ (for which $C_F\to0$ at large time differences, as required there), then shows that $K_F$ and $\mathcal N_F$ satisfy Eq.~\eqref{eq:FDT_bath_discrete} for any $c$: the bath is in equilibrium at the inverse temperature $\beta$.
The FDT of $C_F$ and $G_F$ is an assumption on the solution and is not derived here; the argument shows that it is consistent with the effective process, because the local equilibrium obtained below in turn gives the FDT on the fast time scale.

\paragraph{Static response.}
The static response of the bath, $r_F=\sum_{s\le t}K_F(t,s)$, is the value of $\tilde K_F(\omega)=\RR_A(\tilde G_F(\omega))$ at $\omega=0$, that is, $r_F=\RR_A(\chi_F)=\RR(\chi_F)+c$.

\paragraph{Relation to the fluctuations.}
In a stationary state, Eq.~\eqref{eq:FDT_bath_discrete} also relates $r_F$ to the size of the bath's fluctuations.
With $\tau=t-s$, it reads $K_F(\tau)=\beta[\mathcal N_F(\tau)-\mathcal N_F(\tau+1)]$ for $\tau\ge0$, so the sum over $\tau$ telescopes:
\begin{equation*}
r_F=\sum_{\tau\ge0}K_F(\tau)=\beta\sum_{\tau\ge0}\big[\mathcal N_F(\tau)-\mathcal N_F(\tau+1)\big]=\beta\,\mathcal N_F(0),
\end{equation*}
because $\mathcal N_F(\tau)\to0$ as $\tau\to\infty$.
Since $G(t,t)=0$, $\Gamma_F(t,t)=0$ and $K_F(0)=\kappa_1+c$; hence the retarded part alone sums to $\sum_{s<t}\Gamma_F(t,s)=\beta\,\mathcal N_F(0)-(\kappa_1+c)$.
The relation $r_F=\beta\,\mathcal N_F(0)$, static response equal to $\beta$ times the variance of the fluctuations, is the discrete-time counterpart of the Kramers--Kronig relation used in Appendix~\ref{app:FDT}; it is what makes the bath contribute a factor of the form $\exp(\tfrac\beta2 r_F\sigma^2)$ when it is integrated out.
For the SK model, for example, $\mathcal N_F(0)=(1-q)+c/\beta$, and $\beta\,\mathcal N_F(0)=\chi_F+c=\RR_A(\chi_F)$.

\paragraph{Local equilibrium.}
As in Appendix~\ref{app:FDT} for asynchronous MH, the stationary distribution of the spin at fixed $h$ is obtained by integrating out the bath, which contributes the factor $\exp\{\tfrac\beta2[\RR(\chi_F)+c]\sigma^2\}$.
Because $\sigma^2=1$, this factor does not depend on $\sigma$, and
\begin{equation*}
P(\sigma\,|\,h)\propto e^{\beta h\sigma}\,e^{\frac\beta2[\RR(\chi_F)+c]\sigma^2}\propto e^{\beta h\sigma}.
\end{equation*}
Thus $c$ enters only through a factor independent of $\sigma$, and the self-coupling leaves the local equilibrium unchanged.
For asynchronous MH the bath contains neither the self-coupling nor $\eta$, and its static response is $\RR(\chi_F)-\kappa_1$ instead; the result $P(\sigma|h)\propto e^{\beta h\sigma}$ is the same.

\subsection{Consequences}

\paragraph{Breakdown of the FDT.}
For both dynamics, the stationary equilibrium solution becomes unstable when
$\beta^2\RR'(\beta(1-q))\,\E_z\,\mathrm{sech}^4\beta h>1$, which is the replicon instability condition.
Here $h=\sqrt{q\,\RR'(\beta(1-q))}\,z$, $\E_z$ denotes the average over $z\sim\mathcal N(0,1)$, and $q=\E_z\tanh^2\beta h$ is the plateau value of the replica-symmetric solution.
For the SK model this is the de Almeida--Thouless condition~\cite{AlmeidaThouless1978}.
The derivation is given in Appendix~\ref{app:FDT}.

\paragraph{SK model.}
Under the generalized fluctuation--dissipation relation with a hierarchy of time scales, both dynamics lead to the Parisi equation
$-\partial_q\varphi=\frac{1}{2}[\partial_h^2\varphi+x(q)(\partial_h\varphi)^2]$ with $\varphi(q_{\rm EA},h)=\log2\cosh\beta h$,
where $x(q)$ is fixed by the replicon marginal-stability condition.
For SyncMC this derivation was given in Ref.~\cite{Kabashima2026JSTAT}; the argument above shows that it applies to asynchronous MH without change.
Appendix~\ref{app:Langevin} writes the evolution of the slow components of the local field as Langevin equations, one for each time scale, for a general spectrum.

\paragraph{ROM.}
For SyncMC, the generalized fluctuation--dissipation relation below the dynamical transition was obtained in Ref.~\cite{Kabashima2026PRL}.
Under one-step aging, both dynamics lead to the same plateau value $q$ and fluctuation--dissipation ratio $x$, determined by
\begin{align}
q&=\E_x\big[\tanh^2\beta h\big],\\
1&=\beta^2\RR'\big(\beta(1-q)\big)\,\E_x\big[\mathrm{sech}^4\beta h\big],
\end{align}
where $\E_x$ denotes the average over $h\sim\mathcal N(0,\Sigma_q)$ reweighted by $(2\cosh\beta h)^x$, and
$\Sigma_q=[\RR(\beta(1-q)+\beta xq)-\RR(\beta(1-q))]/(\beta x)$.
The dynamical transition temperature is $T_d\simeq0.1336$ for both dynamics. Table~\ref{tab:ROM} lists the solutions below $T_d$.
The values at $T=0.1$, $q=0.981$ and $x=0.679$, agree with those obtained for SyncMC in Ref.~\cite{Kabashima2026PRL}.
The averages over $h$ were evaluated with adaptive quadrature; a fixed Gauss--Hermite rule with about $10^2$ nodes is not accurate enough at these low temperatures and shifts $x$ in the third digit.

\begin{table}[t]
\caption{Fluctuation--dissipation ratio $x$ and plateau value $q$ for the ROM, common to SyncMC and asynchronous MH. The first row is the dynamical transition.}
\label{tab:ROM}
\begin{ruledtabular}
\begin{tabular}{ccc}
$T$ & $x$ & $q$\\
\hline
$0.1336$ & $1$ & $0.961$\\
$0.12$ & $0.850$ & $0.971$\\
$0.10$ & $0.679$ & $0.981$\\
$0.08$ & $0.539$ & $0.989$\\
$0.06$ & $0.414$ & $0.994$\\
\end{tabular}
\end{ruledtabular}
\end{table}

\subsection{What differs}
\label{sec:differs}

The two dynamics differ in the shape of the fast relaxation $C_F$, $G_F$, which depends on $c$ for SyncMC;
in the time scale on which the plateau is approached; and in how the slow time scales grow with the waiting time.
The slow-time-scale equations are invariant under reparametrization of time~\cite{ChamonEtAl2002} and do not determine this growth. These differences are measured with one MCS of asynchronous MH corresponding to one SyncMC step (Sec.~\ref{sec:time}).

The simulations of Sec.~\ref{sec:results} show how large these differences are.
In the paramagnetic phase, the equilibrium relaxation time is $148$ SyncMC steps against $14.3$ MCS for the ROM at $T=0.3$ with $c=1$,
and $4.3$ steps against $1.3$ MCS for the SK model at $T=1.5$.
In the aging regime, the energy per spin after a quench reaches the same value after $t_{\rm S}$ SyncMC steps and $t_{\rm MH}$ MCS with
$t_{\rm S}/t_{\rm MH}\simeq340$--$470$ for the ROM at $T=0.1$ (slowly increasing with time) and $t_{\rm S}/t_{\rm MH}\simeq15$--$18$ for the SK model at $T=0.5$.
The same correspondence is found independently in the fluctuation--dissipation relation of the ROM: for the same coupling matrices, $T\chi$ at fixed $C$ agrees between the two dynamics at waiting times related in this way (Sec.~\ref{sec:results_rom}).
Thus, per unit time, SyncMC relaxes more slowly than asynchronous MH, by about one order of magnitude in the paramagnetic phase and by more in the aging regime of the ROM,
while the states that the two dynamics visit on slow time scales are described by the same equations.

\paragraph{Critical slowing down near $T_d$.}
Slightly above $T_d$, the equilibrium correlation function decays in two steps: it approaches a plateau near $q_d$, the plateau value at $T_d$, and then leaves it.
With $\epsilon=(T-T_d)/T_d$, the approach follows $C-q_d\propto t^{-a}$, the departure follows $q_d-C\propto t^{b}$, and the relaxation time grows as $\tau\propto\epsilon^{-\gamma}$ with $\gamma=1/(2a)+1/(2b)$.
The exponents are fixed by a single number $\lambda$ through $\Gamma(1-a)^2/\Gamma(1-2a)=\Gamma(1+b)^2/\Gamma(1+2b)=\lambda$, where $\Gamma$ is the gamma function.
We expect $a$, $b$ and $\gamma$ to be the same for the two dynamics, for the following reason.
The exponents follow from the equation for the small deviation $C-q_d$ on times much longer than one but shorter than $\tau$.
On these times the FDT holds, so the response function equals $-\beta\,\partial_tC$ and is small compared with $C-q_d$ divided by the time.
The SyncMC-specific terms then drop out as in Sec.~\ref{sec:slow}: the instantaneous self-coupling leaves the local equilibrium unchanged, and the covariance of $\eta$, a sum of products of response functions, is of higher order in the small deviation.
The remaining terms, built from $\RR$ and the local equilibrium $P(\sigma|h)\propto e^{\beta h\sigma}$, are common to the two dynamics.
In the analysis of Refs.~\cite{CaltagironeEtAl2012PRL,CaltagironeEtAl2012PRB}, $\lambda$ in this regime is given by static quantities of the replica calculation at $T_d$, and Ref.~\cite{CaltagironeEtAl2012PRB} computed it for the ROM.
For the ROM studied here, in which the eigenvalues $\pm1$ of the coupling matrix have equal weights, the method of Ref.~\cite{CaltagironeEtAl2012PRB} gives
\begin{equation}
\lambda=\frac{2\beta_d^3\RR'(z_d)^3\,\E_1\!\big[2m^2(1-m^2)^2\big]}{\RR''(z_d)+2\beta_d^3\RR'(z_d)^3\,\E_1\!\big[(1-m^2)^3\big]}\simeq0.718,
\end{equation}
where $\beta_d=1/T_d$, $z_d=\beta_d(1-q_d)$, $m=\tanh\beta_dh$, and $\E_1$ is the average of Sec.~\ref{sec:slow} with $x=1$ at $T_d$ and $q=q_d$.
This gives $a\simeq0.319$, $b\simeq0.611$ and $\gamma\simeq2.38$; these values should hold for both dynamics.
The dynamics enters only through the microscopic time scale $t_0$ that connects the fast relaxation to the plateau regime, $C-q_d\propto(t/t_0)^{-a}$ and $\tau=t_0\,\epsilon^{-\gamma}$ up to a constant common to both dynamics.
Hence the ratio of the relaxation times of SyncMC (in steps) and asynchronous MH (in MCS) should tend to a constant, $t_0^{\rm S}/t_0^{\rm MH}$, as $T\to T_d$.
We have not checked this numerically; it would require equilibrium simulations close to $T_d$, at sizes large enough that the finite-$N$ rounding of the transition is small.

\section{Numerical methods}
\label{sec:numerics}

\subsection{Solving the effective processes}

Both effective processes are solved by iterating: sample trajectories with given $C$ and $G$, re-estimate $C$ and $G$, and mix.
This Monte Carlo approach to the effective single-spin process follows Eissfeller and Opper~\cite{EissfellerOpper1992}, who applied it to the zero-temperature parallel dynamics of the SK model.
They did not iterate over the whole time window. Because of causality, $C$ and $G$ up to time $t$ fix the process up to time $t+1$, so they advanced one step at a time:
after the spins at time $t$ are sampled, the new row of $C$ is measured, the noise at time $t$ is drawn conditionally on its past values through an extended triangular factor of its covariance, and the new row of the response is obtained.
For the SK model this scheme applies directly to both effective processes and removes the self-consistency iteration.
For a general spectrum, the new rows of $\Gamma_{\rm MH}$, $\Sigma_\phi$ and $\cD$ require the matrix function $\RR$ of matrices that grow by one row and column at each step.
Reference~\cite{Kabashima2026PRL} solved the effective process of SyncMC for the ROM in this way, in a single pass forward in time.
For the comparison in this paper we use the global iteration described here for both dynamics, which treats them with the same code.
For asynchronous MH, the simplest approach is to discretize time with step $\Delta t$ and to replace the jump process by a Bernoulli process with flip probability $\Delta t\,W$.
The number of time points per unit time is then $1/\Delta t$, compared with one for SyncMC,
and memory and the cost of operations on $T\times T$ matrices grow by factors $(1/\Delta t)^2$ and $(1/\Delta t)^3$, respectively.
We use this scheme; since its error is of order $\Delta t$, we combine the solutions for $\Delta t$ and $\Delta t/2$ by Richardson extrapolation.
Appendix~\ref{app:kink} describes an alternative that samples the continuous-time process without discretizing the flip times
and allows the two-time functions to be stored on a coarser grid.

For the ROM, the kernels are computed as follows.
On the time grid, $\cG$ is a strictly lower triangular matrix, because the response is causal; hence $\cG^T=0$ for a $T\times T$ grid, and every power series in $\cG$ has finitely many terms.
The R-transform~\eqref{eq:R_ROM} is the solution of $z\RR^2+\RR-z=0$, and since $\RR(\cG)$ commutes with $\cG$, the matrix $X=\RR(\cG)$ satisfies $\cG X^2+X-\cG=0$.
We solve this equation by Newton's method, starting from $\Gamma_0=0$.
Given the current approximation $\Gamma_k$, we write the solution as $\Gamma_k+\delta$ and expand $\cG(\Gamma_k+\delta)^2+(\Gamma_k+\delta)-\cG=0$ to first order in $\delta$.
Since $\Gamma_k$ and $\delta$ are power series in $\cG$, they commute, and $\delta$ satisfies $(I+2\cG\Gamma_k)\,\delta=-(\cG\Gamma_k^2+\Gamma_k-\cG)$.
The next approximation $\Gamma_{k+1}=\Gamma_k+\delta$ is therefore
\begin{equation}
\Gamma_{k+1}=(I+2\cG\Gamma_k)^{-1}\big(\cG\Gamma_k^2+\cG\big).
\end{equation}
Because $\cG\Gamma_k$ is strictly lower triangular, $I+2\cG\Gamma_k$ is lower triangular with unit diagonal; it is always invertible, and in practice $\Gamma_{k+1}$ is obtained by forward substitution without forming the inverse.
For example, $\Gamma_1=\cG$, and $\Gamma_2=(I+2\cG^2)^{-1}(\cG+\cG^3)=\cG-\cG^3+2\cG^5-4\cG^7+\cdots$ agrees with $\RR(\cG)=\cG-\cG^3+2\cG^5-5\cG^7+\cdots$ up to the term in $\cG^5$.
Each step roughly doubles the order in $\cG$ up to which $\Gamma_k$ is correct, so after about $\log_2T$ steps $\Gamma_k$ equals $\RR(\cG)$ up to rounding errors.
Since $\kappa_1=0$ for the ROM, this is $\Gamma_{\rm MH}$.

The noise covariances are obtained from the same function applied to block matrices.
For $M=\begin{pmatrix}\cG&\cC\\0&\cG^\top\end{pmatrix}$, the upper-right block of $M^n$ is $\sum_{k=0}^{n-1}\cG^k\cC(\cG^\top)^{n-1-k}$.
Therefore $\Sigma_\phi$ in Eq.~\eqref{eq:Sigma_common} is the upper-right block of $\RR(M)=\sum_n\kappa_{n+1}M^n$.
Similarly, for $M'=\begin{pmatrix}\cG&I\\0&\cG^\top\end{pmatrix}$, the upper-right block of $M'^{\,n+1}$ is $\sum_{j=0}^{n}\cG^j(\cG^\top)^{n-j}$,
so $\cD$ in Eq.~\eqref{eq:D} is the upper-right block of $M'\RR_A(M')$.
The matrices $M$ and $M'$ also vanish when raised to the power $2T$, and $\RR(M)$ and $\RR_A(M')=\RR(M')+cI$ are computed by the same Newton iteration.
The coefficient matrices are then block triangular, with diagonal blocks that are lower and upper triangular with unit diagonal.

\subsection{Estimating the response function}

The response function is estimated without applying a field, either from the derivative of the log path probability (score function; for SyncMC see Refs.~\cite{Kabashima2026PRL,Kabashima2026JSTAT}, for asynchronous MH Eq.~\eqref{eq:score_kink} in Appendix~\ref{app:kink}) or from Gaussian integration by parts (Novikov's formula).
The former is unbiased and requires no matrix inversion; the latter requires inverting the noise covariance and is sensitive to its conditioning.
Eissfeller and Opper~\cite{EissfellerOpper1992} used the latter, as a linear system at each time step; for their deterministic rule $\mathrm{sgn}\,h$ the path probability has no derivative with respect to the field, so the score function is not available.

\paragraph{Variance.}
We measure the integrated response $\chi(t,t_w)$, the change of $\langle\sigma_i(t)\rangle$ per unit of a field $\theta_i$ applied from $t_w$ to $t$:
$\chi(t,t_w)=\sum_{s=t_w}^{t-1}G(t,s)$ for SyncMC and $\chi(t,t_w)=\int_{t_w}^{t}G(t,s)\,ds$ for asynchronous MH.
With the score function, $\chi(t,t_w)=\langle\sigma_i(t)\,K_i(t,t_w)\rangle$, where $K_i(t,t_w)$ is the derivative of the log probability of the path of site $i$ between $t_w$ and $t$ with respect to $\theta_i$.
Because $\langle K_i\rangle=0$ and $\sigma_i^2=1$, the variance of one sample of $\sigma_i(t)K_i$ is bounded by $\langle K_i^2\rangle$, which is a sum of independent contributions from successive steps.
For SyncMC and asynchronous MH, respectively,
\begin{align}
T^2\langle K_i^2\rangle&=\sum_{s=t_w}^{t-1}\big\langle1-\tanh^2\beta x_i(s)\big\rangle\le4\,\langle n_i(t,t_w)\rangle,\\
T^2\langle K_i^2\rangle&=4\int_{t_w}^{t}\big\langle W\big(\sigma_i(s),h_i(s)\big)\,\Theta\big(\sigma_i(s)h_i(s)\big)\big\rangle\,ds=4\,\langle n^{\uparrow}_i(t,t_w)\rangle.
\end{align}
Here $T=1/\beta$, $\Theta$ is the unit step function, $n_i(t,t_w)$ is the number of flips of site $i$ between $t_w$ and $t$, and $n_i^\uparrow(t,t_w)$ is the number of those flips that raise the energy.
The inequality for SyncMC follows from $1-\tanh^2\beta x=(1-\sigma\tanh\beta x)(1+\sigma\tanh\beta x)\le2(1-\sigma\tanh\beta x)$, where $(1-\sigma\tanh\beta x)/2$ is the flip probability.
Thus the variance of $T\chi$ per site is at most four times the mean number of flips in the interval.
It grows linearly in $t-t_w$ at high temperatures, but much more slowly in the aging regime, where most spins rarely flip; it does not depend on the time step.
The same bounds hold for the effective processes of the DMFT, with $x_i$ and $h_i$ replaced by the effective fields.

Novikov's formula instead gives $G$ as the correlation of $\sigma$ with $\Sigma^{-1}$ applied to the noise, where $\Sigma$ is the noise covariance on the time grid
($\Sigma_\phi$ for asynchronous MH and $\Sigma_\phi+\beta^{-1}\cD$ for SyncMC).
For asynchronous MH, $\Sigma_\phi$ has a cusp at equal times, like the covariance of an Ornstein--Uhlenbeck process, so its smallest eigenvalues on a grid of step $\Delta t$ are of order $\Delta t$, and the variance of the estimate grows as the grid is refined.
At finite $N$ the formula is not available for asynchronous MH at all, since that dynamics contains no Gaussian noise.

\paragraph{Choice in this paper.}
All response functions in this paper are obtained from the score function, with no field applied.
For finite-$N$ SyncMC we accumulate $K_i(t,t_w)=\sum_{s=t_w}^{t-1}\beta[\sigma_i(s+1)-\tanh\beta x_i(s)]$, which is exact for any $N$.
For finite-$N$ asynchronous MH we use the continuous-time score
$K_i(t,t_w)=\sum_{\text{flips of }i}\partial_h\log W-\int_{t_w}^{t}\partial_hW\big(\sigma_i(s),h_i(s)\big)\,ds$, where the flips are those between $t_w$ and $t$.
The integral is evaluated by the trapezoidal rule from the exact local fields at the ends of each block of $k$ trials of the delayed update (spacing $k/N$ MCS; $k=64$).
The first term is exact, and the error of the second is of order $(k/N)^2$ per block.
In both cases $K_i$ is accumulated in double precision; in single precision the small increments of nearly frozen spins are lost to rounding at large $t-t_w$, which biases $\chi$.
For the DMFT we use the same estimator for SyncMC, and for asynchronous MH in the Bernoulli discretization
$S_l=\Delta t\,e^{-\Delta tW}\partial_hW/(1-e^{-\Delta tW})$ if the spin flips during the interval $l$ and $S_l=-\Delta t\,\partial_hW$ otherwise, with $W$ and $\partial_hW$ evaluated at the start of the interval; then $\Delta t\,G(t_k,t_l)=\langle\sigma(t_k)S_l\rangle$ for $k>l$.
Error bars are standard errors over independent chains and coupling samples.
As checks, both finite-$N$ estimators reproduce the FDT $T\chi=1-C$ within statistical errors in the paramagnetic phase of the SK model ($T=1.5$) and of the ROM ($T=0.3$; Fig.~\ref{fig:fast}).

\subsection{\texorpdfstring{Finite-$N$}{Finite-N} simulations on GPUs}

SyncMC requires two dense matrix--matrix products per step and runs efficiently on GPUs.
For asynchronous MH we use delayed updates:
spins are updated sequentially within a block of $k$ candidate sites using the $k\times k$ submatrix of $J$,
and the local fields of all spins are then updated by one matrix--matrix product.
The dynamics is identical to trial-by-trial MH, and the number of arithmetic operations per MCS is comparable to that of one SyncMC step.
The wall-clock time, however, also depends on how the sequential part is implemented.

Table~\ref{tab:wall} gives the measured wall-clock times on one NVIDIA H200 GPU in single precision, with $32$ independent chains run simultaneously and $k=64$.
In our implementation the $k$ sequential updates within a block are a loop of separate GPU operations written in Python, and their launch overhead, not arithmetic, sets the time per MCS; it is roughly proportional to $N$.
The time per SyncMC step hardly depends on $N$ up to $N=4096$, for the same reason, and grows at $N=8192$, where the matrix products start to dominate.
Increasing the number of chains from $32$ to $128$ only doubles the time per SyncMC step ($0.27$\,ms at $N=4096$).

\begin{table}[t]
\caption{Wall-clock times on one GPU (NVIDIA H200, single precision, $32$ chains, block length $k=64$ for asynchronous MH).
Upper part: time per SyncMC step and per MCS, without measurement, and for $N=4096$ also with the measurement of the two-time functions; the middle and lower parts use the latter.
Middle part: time per equilibrium relaxation time $\tau_e$, defined by $C(t_w+\tau_e,t_w)=1/e$ in equilibrium, in the paramagnetic phase ($N=4096$).
Lower part: time needed to reach the same energy per spin after a quench from a random configuration, in the aging regime ($N=4096$); $t_{\rm MH}$ and $t_{\rm S}$ are the numbers of MCS and SyncMC steps.
The last column is the ratio of the wall-clock times of asynchronous MH and SyncMC.}
\label{tab:wall}
\begin{ruledtabular}
\begin{tabular}{lccc}
 & SyncMC & Asynchronous MH & Ratio\\
\hline
\multicolumn{4}{l}{Time per step or MCS}\\
$N=1024$ & $0.090$\,ms & $115$\,ms & $1.3\times10^3$\\
$N=2048$ & $0.091$\,ms & $230$\,ms & $2.5\times10^3$\\
$N=4096$ & $0.13$\,ms & $0.54$\,s & $4.0\times10^3$\\
$N=8192$ & $0.49$\,ms & $0.92$\,s & $1.9\times10^3$\\
$N=4096$, with measurement & $0.13$\,ms & $0.63$\,s & $4.8\times10^3$\\
\hline
\multicolumn{4}{l}{Paramagnetic phase: $\tau_e$ and time per $\tau_e$}\\
ROM, $T=0.3$, $c=1$ & $148$ steps, $0.020$\,s & $14.3$ MCS, $7.7$\,s & $3.9\times10^2$\\
SK, $T=1.5$ & $4.3$ steps, $0.58$\,ms & $1.32$ MCS, $0.71$\,s & $1.2\times10^3$\\
\hline
\multicolumn{4}{l}{Aging regime: time to reach the energy of MH at $t_{\rm MH}$}\\
ROM, $T=0.1$, $c=1$, $t_{\rm MH}=40$ & $t_{\rm S}=1.34\times10^4$, $1.8$\,s & $25$\,s & $14$\\
ROM, $T=0.1$, $c=1$, $t_{\rm MH}=2560$ & $t_{\rm S}=1.20\times10^6$, $158$\,s & $1600$\,s & $10$\\
SK, $T=0.5$, $t_{\rm MH}=40$ & $t_{\rm S}=608$, $0.08$\,s & $25$\,s & $3.1\times10^2$\\
SK, $T=0.5$, $t_{\rm MH}=2560$ & $t_{\rm S}=4.49\times10^4$, $5.9$\,s & $1600$\,s & $2.7\times10^2$\\
\end{tabular}
\end{ruledtabular}
\end{table}

Per step, SyncMC is three to four orders of magnitude faster than our implementation of asynchronous MH, but it needs more steps to reach the same state (Sec.~\ref{sec:results}).
In the paramagnetic phase, $\tau_e$ is about $10$ times (ROM) and $3$ times (SK) larger in SyncMC steps than in MCS, and the wall-clock time per $\tau_e$ is still shorter for SyncMC by factors of about $400$ (ROM) and $1200$ (SK).
In the aging regime of the ROM, one MCS corresponds to $340$--$470$ SyncMC steps, and the advantage of SyncMC shrinks to a factor of about $10$, decreasing slowly with time;
for the SK model, one MCS corresponds to $15$--$18$ steps, and the factor stays at about $270$.
These ratios refer to our implementation.
If the sequential updates within a block were fused into a single GPU kernel, asynchronous MH would become faster, possibly by more than an order of magnitude, and the advantage of SyncMC in the aging regime of the ROM could disappear.

\section{Numerical results}
\label{sec:results}

All finite-$N$ simulations start from uniformly random configurations at $t=0$ and use the field-free estimators of the response described in Sec.~\ref{sec:numerics}.
Unless stated otherwise, $N=4096$, and averages are taken over four coupling samples with $32$--$128$ independent chains each.
SyncMC uses $c=1$ for the ROM and $c=-\lambda_{\min}(J)\simeq2.0$ (the smallest value that makes $J+cI$ positive semidefinite for each sample) for the SK model.
Asynchronous MH uses delayed updates with blocks of $64$ trials.
Error bars and the uncertainties quoted below are standard errors over coupling samples.

\subsection{Paramagnetic phase: DMFT versus finite \texorpdfstring{$N$}{N}}

\begin{figure}[!htbp]
\centering\includegraphics[width=0.6\textwidth]{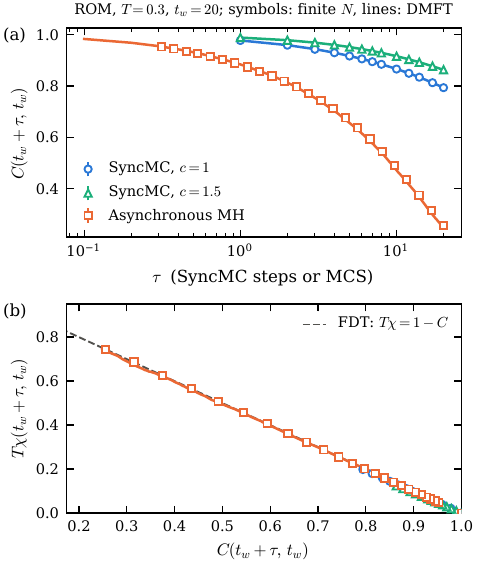}
\caption{ROM at $T=0.3$, $t_w=20$. (a) Correlation function $C(t_w+\tau,t_w)$ versus $\tau$ in SyncMC steps or MCS. Symbols: finite-$N$ simulations ($N=4096$); lines: DMFT.
(b) Integrated response $T\chi$ versus $C$; dashed line: FDT.}
\label{fig:fast}
\end{figure}

Figure~\ref{fig:fast} shows the two-time correlation function of the ROM at $T=0.3$, above $T_d$, at $t_w=20$, together with the solutions of the effective processes.
For SyncMC ($c=1$ and $c=1.5$) the DMFT and the simulations agree within $10^{-3}$ for both $C$ and $T\chi$.
For asynchronous MH, the DMFT solution extrapolated to $\Delta t\to0$ agrees with the simulations within $10^{-2}$ in $C$, down to $C\simeq0.26$.
The finite-$N$ data of asynchronous MH satisfy the FDT, $T\chi=1-C$, within $2\times10^{-3}$.
For SyncMC at this $t_w$ the deviation from the FDT is up to $10^{-2}$; this is because $t_w=20$ steps is short compared with the equilibrium relaxation time of SyncMC ($148$ steps for $c=1$), and the DMFT reproduces this deviation.
Per unit time, SyncMC relaxes much more slowly than asynchronous MH, and more slowly for larger $c$.

\subsection{Aging of the ROM}
\label{sec:results_rom}

\begin{figure}[!htbp]
\includegraphics[width=\textwidth]{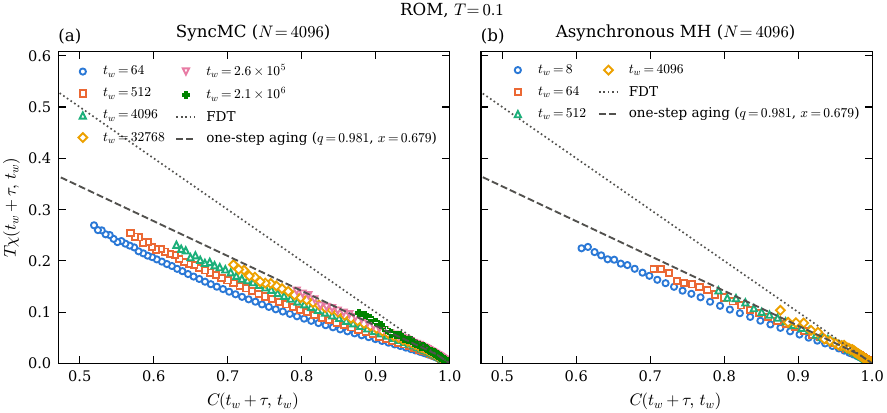}
\caption{Fluctuation--dissipation plots of the ROM at $T=0.1$. (a) SyncMC ($N=4096$); (b) asynchronous MH ($N=4096$). Dotted line: FDT; dashed line: one-step-aging prediction with $q=0.981$ and $x=0.679$, common to both dynamics.}
\label{fig:fdrom}
\end{figure}

Figure~\ref{fig:fdrom} shows the fluctuation--dissipation plots of the ROM at $T=0.1$, below $T_d$, for SyncMC ($N=4096$, $16$ coupling samples, $t_w$ up to $2.1\times10^6$ steps) and for asynchronous MH ($N=4096$, $12$ coupling samples, $t_w$ up to $4096$ MCS); the first $12$ coupling matrices are the same for both dynamics.
For both dynamics, the part with $C>q$ follows the FDT, and the part with $C<q$ approaches, as $t_w$ grows, the common one-step-aging prediction $q=0.981$, $x=0.679$ of Table~\ref{tab:ROM}.
To quantify the approach, Table~\ref{tab:chiC} lists $T\chi$ at fixed values of $C$.
For SyncMC, $T\chi$ at $C=0.93$ increases with $t_w$, reaches the prediction $0.0535$ at $t_w\approx2.6\times10^5$, and stays there within errors up to $t_w=2.1\times10^6$.
For asynchronous MH, $T\chi$ at $C=0.93$ is below the prediction at $t_w=64$ MCS and agrees with it from $t_w=512$ MCS on. Both dynamics thus approach the prediction from below.
The two dynamics take the same values at times related by the energy correspondence of Sec.~\ref{sec:differs}.
Comparing the same coupling matrices sample by sample, the difference of $T\chi$ at $C=0.93$ between asynchronous MH and SyncMC is $-0.0001(5)$ for $t_w=64$ MCS and $3.3\times10^4$ steps, $-0.0020(20)$ for $512$ MCS and $2.6\times10^5$ steps, and $0.003(4)$ for $4096$ MCS and $2.1\times10^6$ steps.
The correspondence of about $10^2$--$10^3$ SyncMC steps per MCS, obtained from the energy, therefore also holds for the fluctuation--dissipation relation.

\begin{table}[t]
\caption{$T\chi(t_w+\tau,t_w)$ at the time $\tau$ at which $C(t_w+\tau,t_w)$ reaches the value in the first row, for the ROM at $T=0.1$. The prediction is $(1-q)+x(q-C)$ with $q=0.981$ and $x=0.679$. Values are obtained from the correlation and response functions averaged over all coupling samples and chains ($16$ samples for SyncMC, $12$ for asynchronous MH). Numbers in parentheses are bootstrap errors in the last digits, from resampling coupling samples and, within each, chains. For SyncMC at $N=8192$ and $t_w=2.1\times10^6$, the lags $\tau\le2^{22}$ are too short for $C$ to reach $0.90$.}
\label{tab:chiC}
\begin{ruledtabular}
\begin{tabular}{llll}
 & $t_w$ & $C=0.93$ & $C=0.90$\\
\hline
Prediction & & $0.0535$ & $0.0738$\\
SyncMC, $N=4096$ & $4096$ & $0.0465(2)$ & $0.0622(3)$\\
 & $32768$ & $0.0507(4)$ & $0.0687(7)$\\
 & $2.6\times10^5$ & $0.0537(14)$ & $0.0713(19)$\\
 & $2.1\times10^6$ & $0.0560(35)$ & $0.0802(62)$\\
SyncMC, $N=2048$ & $2.6\times10^5$ & $0.0569(18)$ & $0.0753(26)$\\
 & $2.1\times10^6$ & $0.0613(43)$ & $0.0783(83)$\\
SyncMC, $N=8192$ & $2.6\times10^5$ & $0.0535(11)$ & $0.0732(19)$\\
 & $2.1\times10^6$ & $0.0519(34)$ & ---\\
Asynchronous MH, $N=4096$ & $64$ & $0.0506(6)$ & $0.0685(8)$\\
 & $512$ & $0.0527(15)$ & $0.0718(20)$\\
 & $4096$ & $0.0557(42)$ & $0.079(7)$\\
Asynchronous MH, $N=2048$ & $64$ & $0.0521(14)$ & $0.0713(21)$\\
 & $512$ & $0.0546(28)$ & $0.0670(49)$\\
 & $4096$ & $0.0545(75)$ & $0.090(11)$\\
\end{tabular}
\end{ruledtabular}
\end{table}

\begin{figure}[!htbp]
\includegraphics[width=\textwidth]{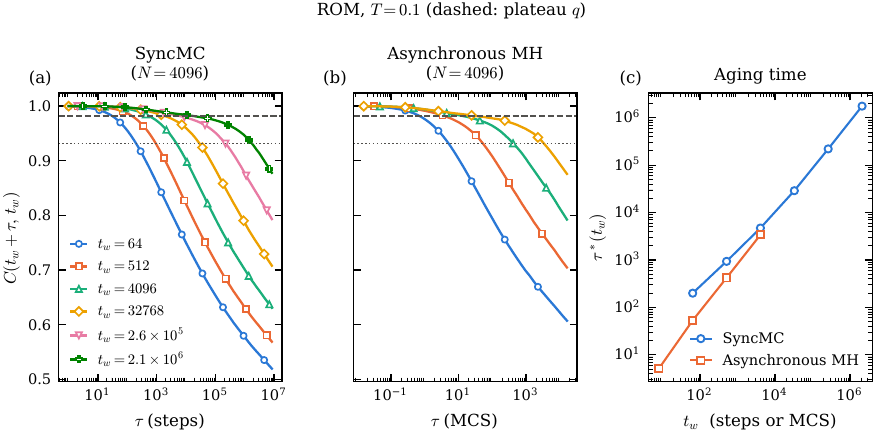}
\caption{Aging of the ROM at $T=0.1$. (a), (b) $C(t_w+\tau,t_w)$ for SyncMC ($N=4096$) and asynchronous MH ($N=4096$). Dashed line: plateau $q$; dotted line: $q-0.05$.
(c) Aging time $\tau^\ast$ at which $C$ reaches $q-0.05$.}
\label{fig:agingrom}
\end{figure}

Figure~\ref{fig:agingrom} shows the correlation functions and the aging time $\tau^\ast(t_w)$, defined by $C(t_w+\tau^\ast,t_w)=q-0.05$.
For both dynamics $\tau^\ast$ grows in proportion to $t_w$ at large $t_w$, with $\tau^\ast/t_w\simeq0.8$--$0.9$ in both cases (SyncMC: $0.88$, $0.84$, $0.84$ for $t_w=3.3\times10^4$, $2.6\times10^5$, $2.1\times10^6$; asynchronous MH: $0.83$, $0.82$, $0.83$ for $t_w=64$, $512$, $4096$).
At small $t_w$, $\tau^\ast/t_w$ is larger for SyncMC ($3.1$ at $t_w=64$), because the fast relaxation of SyncMC itself takes hundreds of steps.

\begin{figure}[!htbp]
\includegraphics[width=\textwidth]{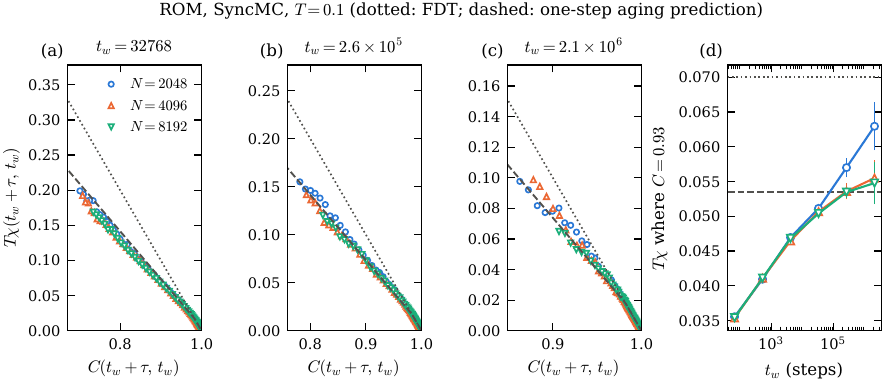}
\caption{Size dependence for SyncMC, ROM at $T=0.1$. (a)--(c) Fluctuation--dissipation plots for $N=2048$, $4096$ and $8192$ ($16$ coupling samples each) at the three largest waiting times. (d) $T\chi$ at $C=0.93$ versus $t_w$, averaged over the values of the individual samples, with standard errors; dashed line: prediction; dotted line: FDT.}
\label{fig:size}
\end{figure}

\paragraph{Finite-size effects.}
Figure~\ref{fig:size} and Table~\ref{tab:chiC} compare SyncMC at $N=2048$, $4096$ and $8192$, with $16$ coupling samples each.
Up to $t_w=3.3\times10^4$ the three sizes agree within errors.
For $N=4096$ and $8192$ the agreement persists, and $T\chi$ at $C=0.93$ stays on the prediction up to $t_w=2.1\times10^6$.
For $N=2048$, $T\chi$ lies above the prediction at $t_w=2.6\times10^5$ and $2.1\times10^6$; the difference from $N=8192$ is about $1.7$ standard errors at each waiting time, and the two values are not independent, since the same samples are followed in time.
A weak finite-size effect at $N=2048$ is therefore possible but not established.
For asynchronous MH, the values for $N=2048$ and $4096$ agree within errors at all waiting times, and both agree with the prediction for $t_w\ge512$ MCS.
Within the times and sizes studied, the data for $N\ge4096$ thus show no finite-size effect for either dynamics.

\subsection{Aging of the SK model}

\begin{figure}[!htbp]
\includegraphics[width=\textwidth]{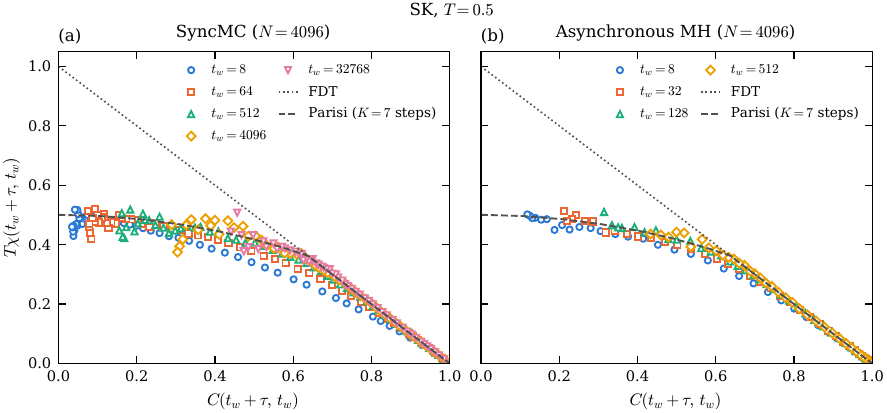}
\caption{Fluctuation--dissipation plots of the SK model at $T=0.5$, $N=4096$. (a) SyncMC; (b) asynchronous MH. Dotted line: FDT; dashed line: prediction from the $K=7$-step Parisi solution, common to both dynamics.}
\label{fig:fdsk}
\end{figure}

Figure~\ref{fig:fdsk} shows the fluctuation--dissipation plots of the SK model at $T=0.5$.
The dashed line is the prediction $T\chi(C)=\int_C^1x(q)\,dq$ from the $K=7$-step Parisi solution.
A $K$-step solution approximates $x(q)$ by a step function with $K$ steps, at overlaps $q_1<\cdots<q_K=q_{\max}$; we obtain it by minimizing the Parisi functional numerically.
As $K$ increases from $1$ to $7$, the minimum of the functional decreases monotonically, and $q_{\max}$ takes the values $0.601$, $0.627$, $0.633$, $0.635$, $0.636$, $0.637$ and $0.637$.
The increments decrease roughly as $K^{-3}$, and extrapolation to $K\to\infty$ gives $q_{\max}\simeq0.638$.
The predicted curve $T\chi(C)$ changes by at most $0.0009$ between $K=6$ and $K=7$, much less than the statistical errors of the data, so the $K=7$ solution is sufficient for the comparison.
For both dynamics the data approach this line as $t_w$ grows.
The mean deviation of $T\chi$ from the prediction for $0.15<C<0.6$ decreases from $-0.051$ at $t_w=8$ to $-0.003$ at $t_w=4096$ for SyncMC,
and from $-0.026$ at $t_w=8$ to within the statistical error ($\simeq0.005$--$0.012$) for $t_w\ge32$ MCS for asynchronous MH.

\section{Discussion}
\label{sec:discussion}

For Ising spin glasses with rotationally invariant couplings, we have written the DMFT of SyncMC and of asynchronous MH in a common form and identified the terms in which they differ.
The differences are an instantaneous self-coupling and a noise whose covariance contains only response functions.
Both act on the fastest time scale, so that the equations for the slow time scales, and with them the predicted plateau value, fluctuation--dissipation ratio and Parisi solution, are the same for the two dynamics.
The underlying reason is that the slow-time-scale equations use the dynamics only through the local equilibrium of a spin in its field and through the fluctuation--dissipation theorem on fast time scales, both of which follow from detailed balance with respect to the Boltzmann distribution.
The update rule enters only through the clock, that is, through how many updates are needed to reach a given state.
The simulations support this picture: at waiting times related by the energy reached after a quench, the two dynamics give the same fluctuation--dissipation relation for the same coupling matrices.

A practical consequence is that slow relaxation and aging of these models can be studied with either dynamics.
The conversion factor between SyncMC steps and MCS depends on the model and on the regime: it is about $10$ in the paramagnetic phase of the ROM, $10^2$--$10^3$ in its aging regime, slowly increasing with time, and $15$--$18$ in the aging regime of the SK model.
Whether SyncMC is faster in wall-clock time therefore depends on this factor and on how efficiently each method is implemented.
In our implementation SyncMC is faster in all cases studied, but an asynchronous MH with the sequential updates fused into a single GPU kernel could reduce or remove this advantage in the aging regime of the ROM.

The conclusions rest on several assumptions and limitations.
The slow-time-scale analysis assumes a separation of time scales, and for the ROM a single aging time scale (one-step aging); the data are consistent with these assumptions but do not prove them.
The DMFT describes the limit $N\to\infty$ at fixed times; in finite systems, processes that cross barriers growing with $N$ eventually stop aging.
Within the sizes and times studied, we found no finite-size effect for $N\ge4096$ and only a weak indication at $N=2048$ for SyncMC.
The prediction that the exponents of critical slowing down near $T_d$ coincide for the two dynamics has not been tested numerically.

The analysis applies to any rotationally invariant ensemble; only the R-transform changes, and ensembles whose R-transform satisfies an algebraic equation can be treated as the ROM was here.
For models with $p$-body interactions, $p\ge3$, the Gaussian auxiliary fields of SyncMC do not apply directly, and a parallel dynamics with the same exact stationary distribution would have to be constructed differently.
The argument above suggests that its slow time scales would again coincide with those of asynchronous dynamics as long as detailed balance holds, but this remains to be shown.

\paragraph*{Data and code availability.}
The code used for the simulations, for the numerical solution of the effective processes and for the figures is available at Zenodo~\cite{ZenodoCode}. The processed data shown in the figures and tables are available from the author upon reasonable request.

\begin{acknowledgments}
Generative-AI tools (Claude Opus 5.5 and Claude Fable 5.1, Anthropic) were used extensively, under the author's direction, for calculations, code, simulations, data analysis, and drafting the text. The author checked the derivations and results, including by comparing theory with simulation, and takes full responsibility for the content of the manuscript.
This work was supported by JSPS KAKENHI Grant Nos.~22H05117 and 26K02981, and JST, ARiSE, Japan Grant Number JPMJAR2633.
\end{acknowledgments}

\appendix
\section{Continuous-time limit of the path weight of asynchronous MH}
\label{app:cont}

We derive Eq.~\eqref{eq:wMHcont} from Eq.~\eqref{eq:wMH}.
Throughout, the history of the field $h(t)$ is given, and $h_k$ denotes its value held fixed on the interval $[t_k,t_{k+1})$, $t_k=k\Delta t$, of the coarse grid; the number of intervals is $T=L/\Delta t$.

\paragraph{Transition probability of one interval.}
Expanding the matrix exponential in Eq.~\eqref{eq:wMH} and using the definition of $\mathbb W$ gives
\begin{equation}
\big[e^{\Delta t\,\mathbb W(h_k)}\big]_{\sigma',\sigma}=
\begin{cases}
1-\Delta t\,W(\sigma,h_k)+O(\Delta t^2), & \sigma'=\sigma,\\
\Delta t\,W(\sigma,h_k)+O(\Delta t^2), & \sigma'=-\sigma.
\end{cases}
\label{eq:oneinterval}
\end{equation}
The first line is the probability that the spin has the same value at the end of the interval as at its beginning, and the second the probability that it has the opposite value.
Trajectories with two or more flips within one interval contribute only to the $O(\Delta t^2)$ terms.

\paragraph{Classification of trajectories on the grid.}
A trajectory on the grid, $(\sigma_0,\sigma_1,\dots,\sigma_T)$ with $\sigma_k=\sigma(t_k)$, is specified by $\sigma_0$ and by the intervals $k_1<k_2<\cdots<k_n$ at whose ends the spin has changed sign, $\sigma_{k_j+1}=-\sigma_{k_j}$.
In Eq.~\eqref{eq:wMH}, each of these $n$ intervals contributes the factor $\Delta t\,W(\sigma_{k_j},h_{k_j})$, where $\sigma_{k_j}$ is the value before the flip,
and each of the remaining $T-n$ intervals contributes $1-\Delta t\,W(\sigma_k,h_k)$. The product of the latter is
\begin{equation}
\prod_{k\notin\{k_1,\dots,k_n\}}\big(1-\Delta t\,W(\sigma_k,h_k)\big)
=\exp\Big(-\Delta t\sum_{k=0}^{T-1}W(\sigma_k,h_k)\Big)\big(1+O(L\,\Delta t)+O(n\,\Delta t)\big).
\end{equation}
Here we used $\log(1-\Delta t\,W)=-\Delta t\,W+O(\Delta t^2)$, summed over at most $T=L/\Delta t$ intervals,
and extended the sum to all $k$, which changes it by at most $n\,\Delta t$ because $W\le1$. Therefore
\begin{equation}
w_{\rm MH}[\sigma\,|\,h]=p_0(\sigma_0)\,(\Delta t)^n\prod_{j=1}^{n}W(\sigma_{k_j},h_{k_j})\;\exp\Big(-\Delta t\sum_{k=0}^{T-1}W(\sigma_k,h_k)\Big)\big(1+O(L\,\Delta t)\big)
\label{eq:wgrid}
\end{equation}
for a trajectory with a fixed number $n$ of flips.

\paragraph{Limit $\Delta t\to0$.}
As $\Delta t\to0$, the field $h_k$ held fixed on each interval converges to $h(t)$, and the sum in the exponent converges to the integral $\int_0^LW(\sigma(t),h(t))\,dt$.
The interval $k_j$ in which the $j$-th flip occurs determines the flip time $\tau_j$ up to $\Delta t$, $\tau_j\in[t_{k_j},t_{k_j+1})$, and $\sigma_{k_j}$ becomes $\sigma(\tau_j^-)$.
The factor $(\Delta t)^n$ is the volume of the set of flip times compatible with the given intervals $k_1,\dots,k_n$.
Dividing Eq.~\eqref{eq:wgrid} by this volume gives a probability density for the flip times,
\begin{equation}
\frac{w_{\rm MH}[\sigma\,|\,h]}{(\Delta t)^n}\;\longrightarrow\;
p_0\big(\sigma(0)\big)\prod_{j=1}^{n}W\big(\sigma(\tau_j^-),h(\tau_j)\big)\exp\Big(-\int_0^LW\big(\sigma(t),h(t)\big)\,dt\Big),
\label{eq:wdensity}
\end{equation}
which is Eq.~\eqref{eq:wMHcont} after writing $\prod_jW=\exp(\sum_j\log W)$.
Accordingly, a sum over trajectories on the grid becomes, in the limit,
\begin{equation}
\sum_{\sigma_0,\dots,\sigma_T}\;\longrightarrow\;\sum_{\sigma(0)=\pm1}\;\sum_{n\ge0}\;\int_{0<\tau_1<\cdots<\tau_n<L}d\tau_1\cdots d\tau_n .
\end{equation}

\paragraph{Normalization.}
For a given field history, the density of Eq.~\eqref{eq:wdensity} integrates to one under this measure.
This can be checked directly: the probability that the spin keeps its value $\sigma$ from $s$ to $t$ is $\exp(-\int_s^tW(\sigma,h(u))\,du)$,
and the density of the first flip at time $\tau$ is $W(\sigma,h(\tau))\exp(-\int_s^\tau W(\sigma,h(u))\,du)$.
Multiplying these factors along a trajectory reproduces Eq.~\eqref{eq:wdensity}, and summing over the number of flips and integrating over their times gives one,
as for any two-state jump process with a time-dependent flip rate.

\section{Saddle point for asynchronous MH}
\label{app:MH}

With $K=-i\Lambda$ and $\langle\sigma(t)\hat h(s)\rangle=iG(t,s)$ at the saddle point,
\begin{equation}
KQ=\begin{pmatrix}\cG^\top&0\\-i\cC&\cG\end{pmatrix},\qquad
(KQ)^n=\begin{pmatrix}(\cG^\top)^n&0\\-i\sum_{k=0}^{n-1}\cG^k\cC(\cG^\top)^{n-1-k}&\cG^n\end{pmatrix}.
\end{equation}
The stationarity condition $\hat Q=K\,\RR(KQ)^\top$ for the conjugate order parameters gives a coupling $-i\hat h^\top\RR(\cG)\sigma$
and a quadratic term $-\frac12\hat h^\top\Sigma_\phi\hat h$ in the single-site weight.
Integrating over $\hat h$ yields Eqs.~\eqref{eq:MH_eff} and \eqref{eq:Sigma_common}.

\section{Saddle point for SyncMC}
\label{app:Sync}

We write out the steps from the generating functional of Sec.~\ref{sec:single} to Eq.~\eqref{eq:Sync_eff}.

\paragraph{Stationarity conditions.}
The exponent of the generating functional is
\begin{equation*}
N\Big[\frac12\Tr\,\mathcal R_A(KQ)-\frac12\Tr\,\hat QQ+\log Z_1[\hat Q]\Big],
\end{equation*}
where $\mathcal R_A$ is the integrated R-transform of the spectrum of $A$, so that $\mathcal R_A'=\RR_A$.
Since $d\,\Tr\,\mathcal R_A(KQ)=\Tr\big[\RR_A(KQ)K\,dQ\big]$, stationarity with respect to $Q$ and $\hat Q$ gives
\begin{equation}
\hat Q=\big[\RR_A(KQ)K\big]^\top=K\,\RR_A(KQ)^\top,\qquad Q=\big\langle\bm y\bm y^\top\big\rangle_1,
\label{eq:statSync}
\end{equation}
where $K$ is symmetric and $\langle\cdots\rangle_1$ denotes the average with the single-site weight in Eq.~\eqref{eq:Z1} ($\epsilon=0$).

\paragraph{Order parameters and $KQ$.}
With $\bm y=(\bm\sigma,\hat{\bm x})$, the blocks of $Q$ are $\langle\sigma(t)\sigma(s)\rangle_1=C(t,s)$, $\langle\sigma(t)\hat x(s)\rangle_1=iG(t,s)$ and $\langle\hat x(t)\hat x(s)\rangle_1=0$.
The second follows because $\theta$ enters Eq.~\eqref{eq:Z1} only through $e^{-i\hat x\theta}$, and the third holds at the saddle point for a normalized weight.
Thus
\begin{equation}
Q=\begin{pmatrix}\cC&i\cG\\i\cG^\top&0\end{pmatrix},\qquad
KQ=\begin{pmatrix}0&-iI_T\\-iI_T&-\beta^{-1}I_T\end{pmatrix}\begin{pmatrix}\cC&i\cG\\i\cG^\top&0\end{pmatrix}
=\begin{pmatrix}\cG^\top&0\\Y_0&\cG\end{pmatrix},
\end{equation}
with $Y_0=-i(\cC+\beta^{-1}\cG^\top)$.

\paragraph{The matrix function $\RR_A(KQ)$.}
For a block lower triangular matrix with diagonal blocks $\cG^\top$ and $\cG$ and lower-left block $Y_0$, induction on $n$ gives
\begin{equation}
(KQ)^n=\begin{pmatrix}(\cG^\top)^n&0\\L_n&\cG^n\end{pmatrix},\qquad L_n=\sum_{k=0}^{n-1}\cG^k\,Y_0\,(\cG^\top)^{n-1-k},
\end{equation}
because $L_{n+1}=Y_0(\cG^\top)^n+\cG L_n$.
With $\RR_A(z)=\sum_{n\ge0}\kappa^A_{n+1}z^n$, where $\kappa^A_1=\kappa_1+c$ and $\kappa^A_{n}=\kappa_n$ for $n\ge2$,
\begin{equation}
\RR_A(KQ)=\begin{pmatrix}\RR_A(\cG^\top)&0\\L&\RR_A(\cG)\end{pmatrix},\qquad L=\sum_{n\ge1}\kappa_{n+1}L_n .
\end{equation}

\paragraph{The conjugate order parameters.}
From Eq.~\eqref{eq:statSync},
\begin{equation}
\hat Q=\begin{pmatrix}0&-iI_T\\-iI_T&-\beta^{-1}I_T\end{pmatrix}\begin{pmatrix}\RR_A(\cG)&L^\top\\0&\RR_A(\cG^\top)\end{pmatrix}
=\begin{pmatrix}0&-i\RR_A(\cG^\top)\\-i\RR_A(\cG)&-iL^\top-\beta^{-1}\RR_A(\cG^\top)\end{pmatrix}.
\end{equation}
Although it does not look so, $\hat Q$ is symmetric, as it must be.
Since $K(QK)^n=(KQ)^nK$ for every $n$, we have $K\RR_A(QK)=\RR_A(KQ)K$; with $K$ and $Q$ symmetric, $\RR_A(KQ)^\top=\RR_A(QK)$, and therefore $\hat Q^\top=\RR_A(KQ)K=K\RR_A(QK)=\hat Q$.
For the lower-right block, symmetry is the identity
\begin{equation}
i\big(L-L^\top\big)=\beta^{-1}\big[\RR_A(\cG^\top)-\RR_A(\cG)\big],
\label{eq:Lantisym}
\end{equation}
which can also be checked directly from the definition of $L$.
With $iY_0=\cC+\beta^{-1}\cG^\top$ and $iY_0^\top=\cC+\beta^{-1}\cG$, the terms with $\cC$ cancel in $i(L-L^\top)$ because $\cC$ is symmetric, and
\begin{equation*}
i\big(L-L^\top\big)=\beta^{-1}\sum_{n\ge1}\kappa_{n+1}\Big[\sum_{k=0}^{n-1}\cG^k(\cG^\top)^{n-k}-\sum_{k=1}^{n}\cG^k(\cG^\top)^{n-k}\Big]
=\beta^{-1}\sum_{n\ge1}\kappa_{n+1}\big[(\cG^\top)^n-\cG^n\big],
\end{equation*}
which is Eq.~\eqref{eq:Lantisym}; the constant $\kappa^A_1=\kappa_1+c$ cancels on the right-hand side.
The term $\frac12\bm y^\top\hat Q\bm y$ in Eq.~\eqref{eq:Z1} therefore equals
\begin{equation}
\frac12\bm y^\top\hat Q\bm y=-i\,\hat{\bm x}^\top\RR_A(\cG)\,\bm\sigma-\frac12\,\hat{\bm x}^\top\mathcal N\,\hat{\bm x},\qquad
\mathcal N=\frac12\Big[i\big(L+L^\top\big)+\beta^{-1}\big(\RR_A(\cG)+\RR_A(\cG^\top)\big)\Big],
\end{equation}
where we used $\bm\sigma^\top\RR_A(\cG^\top)\hat{\bm x}=\hat{\bm x}^\top\RR_A(\cG)\bm\sigma$.
By Eq.~\eqref{eq:Lantisym}, the lower-right block $-iL^\top-\beta^{-1}\RR_A(\cG^\top)$ of $\hat Q$ equals its own symmetric part, which is $-\mathcal N$; writing it in this symmetric form is an identity, not an approximation.

\paragraph{The noise covariance.}
Inserting $Y_0$,
\begin{equation}
iL=\sum_{n\ge1}\kappa_{n+1}\sum_{k=0}^{n-1}\cG^k\cC(\cG^\top)^{n-1-k}+\beta^{-1}\sum_{n\ge1}\kappa_{n+1}\sum_{k=0}^{n-1}\cG^k(\cG^\top)^{n-k}.
\end{equation}
The first sum is $\Sigma_\phi$ of Eq.~\eqref{eq:Sigma_common}; it is symmetric because $\cC$ is.
For the second, the symmetric part is
\begin{equation*}
\frac12\sum_{k=0}^{n-1}\big[\cG^k(\cG^\top)^{n-k}+\cG^{n-k}(\cG^\top)^{k}\big],
\end{equation*}
and adding the term $\frac12(\cG^n+\cG^{\top n})$ from $\RR_A(\cG)+\RR_A(\cG^\top)$ gives
\begin{equation}
\frac12\sum_{k=0}^{n-1}\cG^k(\cG^\top)^{n-k}+\frac12\sum_{j=1}^{n}\cG^j(\cG^\top)^{n-j}+\frac12\big(\cG^n+\cG^{\top n}\big)
=\sum_{j=0}^{n}\cG^j(\cG^\top)^{n-j}\qquad(n\ge1).
\end{equation}
The term $n=0$ of $\RR_A(\cG)+\RR_A(\cG^\top)$ contributes $\kappa^A_1I_T$, which is the term $n=0$ of the same sum. Altogether,
\begin{equation}
\mathcal N=\Sigma_\phi+\beta^{-1}\cD,\qquad \cD=\sum_{n\ge0}\kappa^A_{n+1}\sum_{j=0}^{n}\cG^j(\cG^\top)^{n-j},
\end{equation}
which is Eq.~\eqref{eq:D}.

\paragraph{The effective process.}
The single-site generating functional becomes
\begin{equation}
Z_1=\sum_{\sigma}\int\prod_t\frac{dx_t\,d\hat x_t}{2\pi}\;w_{\rm S}[\sigma\,|\,x]\,
\exp\Big(i\,\hat{\bm x}^\top\big(\bm x-\bm\theta-\RR_A(\cG)\bm\sigma\big)-\frac12\,\hat{\bm x}^\top\mathcal N\hat{\bm x}\Big).
\end{equation}
The Gaussian integral over $\hat{\bm x}$ gives the probability density of $\bm x=\RR_A(\cG)\bm\sigma+\bm\theta+\bm\xi$ with $\bm\xi\sim\mathcal N(0,\mathcal N)$.
Since $\RR_A(\cG)=(\kappa_1+c)I_T+\sum_{n\ge1}\kappa_{n+1}\cG^n=(\kappa_1+c)I_T+\Gamma_{\rm MH}$, and $\bm\xi$ can be written as the sum of independent Gaussian processes $\phi$ and $\eta$ with covariances $\Sigma_\phi$ and $\beta^{-1}\cD$,
the field is that of Eq.~\eqref{eq:Sync_eff}. Together with the weight $w_{\rm S}$, which draws $\sigma(t+1)$ by the heat-bath rule in $x(t)$, this is the effective process of SyncMC.

\section{Fourier representation of the common noise}
\label{app:Fourier}

We derive Eqs.~\eqref{eq:SigmaFourier} and \eqref{eq:FDTpreserve}, and then the discrete-time relations \eqref{eq:FDT_discrete}--\eqref{eq:FDT_bath_discrete} of SyncMC.
The first part is written for continuous time (asynchronous MH);
for SyncMC the same steps hold with integrals over time replaced by sums over steps and $\omega$ restricted to $-\pi<\omega\le\pi$.

\paragraph{Operators that depend only on time differences.}
We consider a stationary state, in which the time origin has been moved to $-\infty$ and all two-time functions depend only on $\tau=t-s$.
An operator $\mathcal A$ with such a kernel acts on a function $f$ as a convolution, $(\mathcal Af)(t)=\int ds\,A(t-s)f(s)$.
With the Fourier transform $\tilde f(\omega)=\int d\tau\,e^{i\omega\tau}f(\tau)$, the following three rules hold.
\begin{enumerate}
\item The product of two such operators is again such an operator, and its Fourier transform is the product of the Fourier transforms:
$\widetilde{(\mathcal A\mathcal B)}(\omega)=\tilde A(\omega)\tilde B(\omega)$. In particular, $\widetilde{(\cG^k)}(\omega)=\tilde G(\omega)^k$.
\item The transposed operator has kernel $A^\top(t,s)=A(s,t)=A(-\tau)$. For a real kernel its Fourier transform is
$\int d\tau\,e^{i\omega\tau}A(-\tau)=\tilde A(-\omega)=\tilde A(\omega)^*$. In particular, $\widetilde{(\cG^\top)}(\omega)=\tilde G(\omega)^*$.
\item Since the Fourier transforms are numbers for each $\omega$, the order of the factors no longer matters.
\end{enumerate}

\paragraph{Derivation of Eq.~\eqref{eq:SigmaFourier}.}
Applying these rules to each term of Eq.~\eqref{eq:Sigma_common} gives
\begin{equation}
\tilde\Sigma_\phi(\omega)=\sum_{n\ge1}\kappa_{n+1}\sum_{k=0}^{n-1}\tilde G(\omega)^k\,\tilde C(\omega)\,\big(\tilde G(\omega)^*\big)^{n-1-k}
=\tilde C(\omega)\sum_{n\ge1}\kappa_{n+1}\sum_{k=0}^{n-1}\tilde G(\omega)^k\big(\tilde G(\omega)^*\big)^{n-1-k}.
\end{equation}
The inner sum is a finite geometric sum,
\begin{equation}
\sum_{k=0}^{n-1}a^kb^{n-1-k}=\frac{a^n-b^n}{a-b}\qquad(a\neq b),
\end{equation}
with $a=\tilde G(\omega)$ and $b=\tilde G(\omega)^*$. Hence
\begin{equation}
\tilde\Sigma_\phi(\omega)=\tilde C(\omega)\,\frac{\sum_{n\ge1}\kappa_{n+1}\big[\tilde G(\omega)^n-\big(\tilde G(\omega)^*\big)^n\big]}{\tilde G(\omega)-\tilde G(\omega)^*}
=\tilde C(\omega)\,\frac{\RR\big(\tilde G(\omega)\big)-\RR\big(\tilde G(\omega)^*\big)}{\tilde G(\omega)-\tilde G(\omega)^*},
\end{equation}
where we added the $n=0$ term $\kappa_1$, which cancels in the difference, to obtain $\RR(z)=\sum_{n\ge0}\kappa_{n+1}z^n$.
At frequencies where $\tilde G(\omega)$ is real, such as $\omega=0$, the ratio is understood as its limit $\RR'(\tilde G(\omega))$.
Since $C$ is real and symmetric, $C(-\tau)=C(\tau)$, its Fourier transform $\tilde C(\omega)$ is real; since the coefficients $\kappa_n$ are real, $\RR(z^*)=\RR(z)^*$,
so the ratio is real as well, and $\tilde\Sigma_\phi(\omega)$ is real, as a covariance of a stationary process must be.

\paragraph{The FDT in Fourier space.}
Suppose $G(\tau)=-\beta_{\rm e}C'(\tau)\Theta(\tau)$, where $\beta_{\rm e}$ is a general inverse temperature ($\beta_{\rm e}=\beta$ in equilibrium, $\beta_{\rm e}=\beta x$ on a slow time scale with fluctuation--dissipation ratio $x$), and that $C(\tau)\to0$ as $\tau\to\infty$
(a constant part of $C$ is treated separately, as in Appendix~\ref{app:FDT}). Integrating by parts,
\begin{equation}
\tilde G(\omega)=-\beta_{\rm e}\int_0^\infty d\tau\,e^{i\omega\tau}C'(\tau)=\beta_{\rm e}\Big[C(0)+i\omega\int_0^\infty d\tau\,e^{i\omega\tau}C(\tau)\Big].
\end{equation}
Taking the imaginary part and using $C(-\tau)=C(\tau)$, so that $\int_0^\infty d\tau\,\cos(\omega\tau)\,C(\tau)=\tilde C(\omega)/2$, gives
\begin{equation}
\mathrm{Im}\,\tilde G(\omega)=\frac{\beta_{\rm e}\omega}{2}\,\tilde C(\omega),
\qquad\text{that is,}\qquad
\tilde C(\omega)=\frac{2}{\beta_{\rm e}\omega}\,\mathrm{Im}\,\tilde G(\omega).
\end{equation}

\paragraph{Derivation of Eq.~\eqref{eq:FDTpreserve}.}
Using $\tilde G-\tilde G^*=2i\,\mathrm{Im}\,\tilde G$ and $\RR(\tilde G)-\RR(\tilde G^*)=\RR(\tilde G)-\RR(\tilde G)^*=2i\,\mathrm{Im}\,\RR(\tilde G)$ in Eq.~\eqref{eq:SigmaFourier}, we obtain
\begin{equation}
\tilde\Sigma_\phi(\omega)=\tilde C(\omega)\,\frac{\mathrm{Im}\,\RR\big(\tilde G(\omega)\big)}{\mathrm{Im}\,\tilde G(\omega)}
=\frac{2}{\beta_{\rm e}\omega}\,\mathrm{Im}\,\RR\big(\tilde G(\omega)\big),
\end{equation}
where the FDT was used in the second step. The Fourier transform of the retarded self-interaction is $\tilde\Gamma_{\rm MH}(\omega)=\RR(\tilde G(\omega))-\kappa_1$, by rules 1 and 3,
and $\mathrm{Im}\,\tilde\Gamma_{\rm MH}=\mathrm{Im}\,\RR(\tilde G)$. Hence $\tilde\Sigma_\phi(\omega)=(2/\beta_{\rm e}\omega)\,\mathrm{Im}\,\tilde\Gamma_{\rm MH}(\omega)$:
the noise $\phi$ and the retarded self-interaction $\Gamma_{\rm MH}$ are related in the same way as $C$ and $G$, that is, they satisfy the FDT at the same inverse temperature $\beta_{\rm e}$.

\paragraph{Discrete-time FDT of SyncMC: derivation of Eq.~\eqref{eq:FDT_discrete}.}
In SyncMC the score of the field $\theta_i(s)$ is $\beta[\sigma_i(s+1)-\tanh\beta x_i(s)]$, so that
$G(t,s)=\beta\langle\sigma_i(t)\sigma_i(s+1)\rangle-\beta\langle\sigma_i(t)\tanh\beta x_i(s)\rangle$ for $t\ge s+1$.
In equilibrium, $(\bm\sigma(s),\bm x(s))$ is distributed according to $p(\bm\sigma,\bm x)$ of Sec.~\ref{sec:model}, under which the conditional mean of $\sigma_i(s)$ given $\bm x(s)$ is $\tanh\beta x_i(s)$.
Moreover, $\sigma_i(t)$ for $t\ge s+1$ depends on $\bm\sigma(s)$ only through $\bm x(s)$. Hence $\langle\sigma_i(t)\tanh\beta x_i(s)\rangle=\langle\sigma_i(t)\sigma_i(s)\rangle$, which gives Eq.~\eqref{eq:FDT_discrete}.

\paragraph{Fourier form.}
Let $S(\omega)=\sum_{\tau\ge0}e^{i\omega\tau}C(\tau)$, so that $\tilde C=2\,\mathrm{Re}\,S-C(0)$. Equation~\eqref{eq:FDT_discrete} gives
$\tilde G=\beta\sum_{\tau\ge1}e^{i\omega\tau}[C(\tau-1)-C(\tau)]=\beta[C(0)+(e^{i\omega}-1)S]$.
Multiplying by $e^{-i\omega/2}$ and using $e^{i\omega/2}-e^{-i\omega/2}=2i\sin(\omega/2)$,
$\mathrm{Im}[e^{-i\omega/2}\tilde G]=\beta\sin(\omega/2)[2\,\mathrm{Re}\,S-C(0)]=\beta\sin(\omega/2)\,\tilde C$, which is Eq.~\eqref{eq:FDT_discrete_F}.
In the same way for the bath: with $\tau=t-s$, Eq.~\eqref{eq:FDT_bath_discrete} reads $K(\tau)=\beta[\mathcal N(\tau)-\mathcal N(\tau+1)]$ for $\tau\ge0$, because $\mathcal N(t,s-1)=\mathcal N(\tau+1)$, and $K(\tau)=0$ for $\tau<0$.
Let $S_{\mathcal N}(\omega)=\sum_{\tau\ge0}e^{i\omega\tau}\mathcal N(\tau)$, so that $\tilde{\mathcal N}=2\,\mathrm{Re}\,S_{\mathcal N}-\mathcal N(0)$ since $\mathcal N$ is symmetric.
Then $\tilde K=\beta[(1-e^{-i\omega})S_{\mathcal N}+e^{-i\omega}\mathcal N(0)]$, and multiplying by $e^{i\omega/2}$,
$\mathrm{Im}[e^{i\omega/2}\tilde K]=\beta\sin(\omega/2)[2\,\mathrm{Re}\,S_{\mathcal N}-\mathcal N(0)]=\beta\sin(\omega/2)\,\tilde{\mathcal N}$, which is the Fourier form stated after Eq.~\eqref{eq:FDT_bath_discrete}.
Conversely, since $K$ vanishes for $\tau<0$, $\mathrm{Im}[e^{i\omega/2}\tilde K(\omega)]=\sum_{\tau\ge0}K(\tau)\sin\big(\omega(\tau+\tfrac12)\big)$ determines $K(\tau)$ for every $\tau\ge0$, so the two forms are equivalent.

\paragraph{Derivation of Eq.~\eqref{eq:FDT_bath_discrete}.}
By the Fourier form above, Eq.~\eqref{eq:FDT_bath_discrete}, that is $K(\tau)=\beta[\mathcal N(\tau)-\mathcal N(\tau+1)]$, is equivalent to
\begin{equation}
\beta\sin\frac\omega2\,\tilde{\mathcal N}(\omega)=\mathrm{Im}\big[e^{i\omega/2}\tilde K(\omega)\big],
\label{eq:bathF}
\end{equation}
so it suffices to prove Eq.~\eqref{eq:bathF}.
We first express both sides through $\tilde G$.
The response of the bath is the operator $K=(\kappa_1+c)I+\Gamma_{\rm MH}=\RR_A(\cG)$, so by rules 1 and 3 its Fourier transform is $\tilde K(\omega)=\RR_A(\tilde G(\omega))$.
The covariance of the bath is $\mathcal N=\Sigma_\phi+\beta^{-1}\cD$, so $\tilde{\mathcal N}=\tilde\Sigma_\phi+\beta^{-1}\tilde\cD$.
Here $\tilde\Sigma_\phi$ is given by Eq.~\eqref{eq:SigmaFourier}, and Eq.~\eqref{eq:D} gives
$\tilde\cD=\sum_{n\ge0}\kappa^A_{n+1}\sum_{j=0}^{n}z^jw^{n-j}=[z\RR_A(z)-w\RR_A(w)]/(z-w)$, with $z=\tilde G(\omega)$ and $w=\tilde G(\omega)^*$.
Write $a=e^{i\omega/2}$, so that $\sin(\omega/2)=(a-a^{-1})/(2i)$, and use Eq.~\eqref{eq:FDT_discrete_F} in the form $\beta\sin(\omega/2)\,\tilde C=(a^{-1}z-aw)/(2i)$. Since $\RR(z)-\RR(w)=\RR_A(z)-\RR_A(w)$,
\begin{align}
2i(z-w)\beta\sin\frac\omega2\Big[\tilde\Sigma_\phi+\frac{\tilde\cD}{\beta}\Big]
&=(a^{-1}z-aw)\big[\RR_A(z)-\RR_A(w)\big]+(a-a^{-1})\big[z\RR_A(z)-w\RR_A(w)\big]\nonumber\\
&=(z-w)\big[a\,\RR_A(z)-a^{-1}\RR_A(w)\big].
\end{align}
Because the coefficients of $\RR_A$ are real, $\RR_A(w)=\RR_A(z)^*$, and with $a^{-1}=a^*$ the right-hand side divided by $2i(z-w)$ is $\mathrm{Im}[a\,\RR_A(z)]$.
The left-hand side divided by $2i(z-w)$ is $\beta\sin(\omega/2)\,\tilde{\mathcal N}$, and $a\,\RR_A(z)=e^{i\omega/2}\tilde K(\omega)$. Hence
$\beta\sin(\omega/2)\,\tilde{\mathcal N}(\omega)=\mathrm{Im}[e^{i\omega/2}\tilde K(\omega)]$, which is Eq.~\eqref{eq:bathF}.
By the Fourier form above, this is $K(\tau)=\beta[\mathcal N(\tau)-\mathcal N(\tau+1)]$ for $\tau\ge0$, that is, Eq.~\eqref{eq:FDT_bath_discrete}. The cancellation in the second line requires both terms; with $\tilde\Sigma_\phi$ alone the relation does not hold, as we have also confirmed numerically for the ROM.

\section{Breakdown of the fluctuation--dissipation theorem}
\label{app:FDT}

We derive the instability condition stated in Sec.~\ref{sec:slow} from the effective process of asynchronous MH, Eqs.~\eqref{eq:MH_eff} and \eqref{eq:Sigma_common}, with $\theta=0$.
By the argument of Sec.~\ref{sec:slow}, the same condition holds for SyncMC.

\paragraph{Equilibrium solution.}
We assume that $C$ and $G$ depend only on $\tau=t-s$ and satisfy the FDT at the inverse temperature $\beta$ of the dynamics, $G(\tau)=-\beta C'(\tau)\Theta(\tau)$.
Let $q=\lim_{\tau\to\infty}C(\tau)$ and $c(\tau)=C(\tau)-q$. The integrated response is
\begin{equation}
\chi\equiv\int_0^\infty G(\tau)\,d\tau=\tilde G(0)=\beta(1-q).
\end{equation}
Because $\tilde C(\omega)=2\pi q\,\delta(\omega)+\tilde c(\omega)$, Eq.~\eqref{eq:SigmaFourier} splits the noise into two independent parts,
\begin{equation}
\phi(t)=\sqrt{q\,\RR'(\chi)}\;z+\tilde\phi(t),\qquad z\sim\mathcal N(0,1),
\end{equation}
where the first part is constant in time, with variance obtained from the ratio in Eq.~\eqref{eq:SigmaFourier} at $\omega=0$, where $\tilde G(0)=\chi$ is real.
The second part $\tilde\phi$ has covariance $\tilde c(\omega)\,[\RR(\tilde G)-\RR(\tilde G^*)]/(\tilde G-\tilde G^*)$, which equals $(2/\beta\omega)\,\mathrm{Im}\,\RR(\tilde G(\omega))$ by the FDT for $c$ and $G$.
Hence the field $\xi(t)=\tilde\phi(t)+\int_0^t\Gamma_{\rm MH}(t,s)\sigma(s)\,ds$ acts on the spin as a Gaussian bath in equilibrium at inverse temperature $\beta$, coupled linearly to $\sigma$ through the energy $-\sigma\xi$.
Its static response is $\tilde\Gamma_{\rm MH}(0)=\RR(\chi)-\kappa_1$, and, by the Kramers--Kronig relation, the variance of its free fluctuations is $[\RR(\chi)-\kappa_1]/\beta$, in agreement with the FDT.
The flip rate $W$ satisfies detailed balance for the spin together with the bath, so the stationary distribution of the spin at fixed $z$ is obtained by integrating out the bath:
the bath contributes a factor $\exp\{\beta[\RR(\chi)-\kappa_1]\sigma^2/2\}$, which is a constant because $\sigma^2=1$. Therefore
\begin{equation}
P(\sigma\,|\,z)=\frac{e^{\beta h_0\sigma}}{2\cosh\beta h_0},\qquad h_0=\sqrt{q\,\RR'\big(\beta(1-q)\big)}\;z,
\end{equation}
and the limit $\tau\to\infty$ of $C(\tau)$ gives
\begin{equation}
q=\E_z\tanh^2\beta h_0,
\label{eq:qRS}
\end{equation}
where $\E_z$ denotes the average over $z$. For the SK model, $\RR'=1$ and Eq.~\eqref{eq:qRS} is the replica-symmetric equation.

\paragraph{Stability.}
We now ask whether $C$ can approach $q$ through a small, slowly decaying tail.
Suppose that $C(\tau)$ contains, in addition to the equilibrium solution, a small component $c_{\rm s}(\tau)$ that varies on a time scale much longer than the fast relaxation.
At frequencies of order of the inverse of this time scale, $\tilde G(\omega)\simeq\chi$ up to corrections of order $c_{\rm s}$,
so Eq.~\eqref{eq:SigmaFourier} gives a slow component of the noise, $\delta\phi_{\rm s}(t)$, with covariance
\begin{equation}
\langle\delta\phi_{\rm s}(t)\,\delta\phi_{\rm s}(s)\rangle=\RR'\big(\beta(1-q)\big)\,c_{\rm s}(t-s)+O(c_{\rm s}^2).
\end{equation}
The slow part of the retarded self-interaction is of order $c_{\rm s}$ and is negligible compared with $\delta\phi_{\rm s}=O(c_{\rm s}^{1/2})$.
Because $\delta\phi_{\rm s}$ changes slowly, the spin and the fast bath remain in the equilibrium described above, with $h_0$ replaced by $h_0+\delta\phi_{\rm s}$.
The local magnetization therefore follows adiabatically,
\begin{equation}
\delta m(t)=\beta\,\mathrm{sech}^2(\beta h_0)\,\delta\phi_{\rm s}(t).
\end{equation}
The slow component of the correlation function is $c_{\rm s}(t-s)=\E_z\langle\delta m(t)\,\delta m(s)\rangle$, where $\langle\cdots\rangle$ is the average over $\delta\phi_{\rm s}$. This gives
\begin{equation}
c_{\rm s}(\tau)=\lambda\,c_{\rm s}(\tau)+o(c_{\rm s}),\qquad
\lambda=\beta^2\,\RR'\big(\beta(1-q)\big)\,\E_z\,\mathrm{sech}^4\beta h_0 .
\label{eq:lambda}
\end{equation}
If $\lambda<1$, the only solution is $c_{\rm s}=0$: the correlation function reaches $q$ in a finite time and the equilibrium solution is stable.
At $\lambda=1$ the relaxation time diverges, and for $\lambda>1$ slow fluctuations are amplified, so that the equilibrium solution, and with it the FDT on all time scales, can no longer hold.
This gives the condition $\lambda>1$ stated in Sec.~\ref{sec:slow}. For the SK model, $\RR'=1$ and $\lambda=\beta^2\E_z\,\mathrm{sech}^4\beta h_0$, the de Almeida--Thouless condition~\cite{AlmeidaThouless1978};
in zero external field $q=0$ and the condition reduces to $\beta>1$.

\section{Langevin equations for the slow components of the local field}
\label{app:Langevin}

We describe how the slow part of the local field evolves once the time scales are separated, following Sec.~7 of Ref.~\cite{Kabashima2026JSTAT} for the SK model.
The construction carries over to any rotationally invariant ensemble.
The one change is that the variance of the noise and the strength of the retarded self-interaction of each time scale
are set by the R-transform evaluated at the integrated response of the faster time scales, rather than directly by $C$ and $G$.
The argument uses only the parts of the effective process that are common to the two dynamics (Sec.~\ref{sec:slow}),
so it applies to asynchronous MH and to SyncMC alike; time is measured in MCS or in SyncMC steps.

\paragraph{Assumptions.}
We use the two assumptions of Sec.~\ref{sec:slow} in the following form.
\begin{enumerate}
\item[(i)] \emph{Separation of time scales.} The correlation function decays on widely separated time scales $1\sim\tau_K\ll\tau_{K-1}\ll\cdots\ll\tau_0$,
\begin{equation}
C(t,s)=q_0+\sum_{k=0}^{K}a_k\,c_k\Big(\frac{t-s}{\tau_k}\Big),\qquad a_k=q_{k+1}-q_k,\qquad q_0\le q_1\le\cdots\le q_{K+1}=1,
\label{eq:Cbands}
\end{equation}
where each $c_k$ decreases from $c_k(0)=1$ to $c_k(\infty)=0$. When $t-s$ passes $\tau_k$, $C$ drops from $q_{k+1}$ to $q_k$.
We call the part of the dynamics on the time scale $\tau_k$ band $k$; band $K$ is the fastest.
\item[(ii)] \emph{Fluctuation--dissipation ratio of each band.} The response of band $k$ satisfies $G=\beta x_k\,\partial_sC$ with the part of $C$ belonging to band $k$,
with $0\le x_0\le\cdots\le x_K=1$. For the fastest band this is the FDT at the temperature of the dynamics.
\end{enumerate}

\paragraph{Integrated responses.}
The response integrated over band $k$ is $\beta x_ka_k$. We write the response integrated over band $k$ and all faster bands as
\begin{equation}
\chi_k=\beta\sum_{j=k}^{K}x_ja_j,\qquad \chi_{K+1}=0 .
\label{eq:chik}
\end{equation}
At time separations of order $\tau_k$, the faster bands $j>k$ have already responded completely and act as an instantaneous response of total weight $\chi_{k+1}$,
while the slower bands $j<k$ have not yet responded. In Fourier space at frequencies $\omega$ of order $1/\tau_k$, this means $\tilde G(\omega)\simeq\chi_{k+1}+\tilde g_k(\omega)$,
where $\tilde g_k$ is the Fourier transform of the response of band $k$, with $\tilde g_k(0)=\beta x_ka_k$.

\paragraph{Noise and retarded self-interaction of band $k$.}
By the rules of Appendix~\ref{app:Fourier}, the retarded self-interaction acting on band $k$ has the Fourier transform
\begin{equation}
\tilde\Gamma_k(\omega)=\RR\big(\chi_{k+1}+\tilde g_k(\omega)\big)-\RR(\chi_{k+1}),
\label{eq:Gammak}
\end{equation}
and, since $C$ and $G$ of band $k$ satisfy the FDT at inverse temperature $\beta x_k$, Eq.~\eqref{eq:FDTpreserve} with $\beta_{\rm e}=\beta x_k$ gives the covariance of the noise of band $k$:
\begin{equation}
\tilde\Sigma_k(\omega)=\frac{2}{\beta x_k\,\omega}\,\mathrm{Im}\,\tilde\Gamma_k(\omega).
\label{eq:Sigmak}
\end{equation}
Thus the noise $\phi_k$ and the retarded self-interaction $\Gamma_k$ of band $k$ have the same time dependence and satisfy the FDT at the effective temperature $T/x_k$.
Their overall sizes follow from the values at $\omega=0$ and, through the Kramers--Kronig relation, at $\omega\to\infty$:
\begin{equation}
w_k\equiv\int_0^\infty\Gamma_k(\tau)\,d\tau=\RR(\chi_k)-\RR(\chi_{k+1}),\qquad
v_k\equiv\Sigma_k(0)=\frac{w_k}{\beta x_k}=\frac{\RR(\chi_k)-\RR(\chi_{k+1})}{\beta x_k}.
\label{eq:wkvk}
\end{equation}
Here $w_k$ is the static response of the noise source of band $k$ to the spin, and $v_k$ is the variance of $\phi_k$.
The constant part $q_0$ of $C$ gives, in addition, a static Gaussian field $z_{-1}$ of variance $v_{-1}=q_0\,\RR'(\chi_0)$, from the ratio in Eq.~\eqref{eq:SigmaFourier} at $\omega=0$.
For the SK model, $\RR(z)=z$, and Eq.~\eqref{eq:wkvk} gives $w_k=\beta x_ka_k$ and $v_k=a_k$: the noise of band $k$ has the variance of the drop of $C$ and the same time dependence $c_k$,
as in Ref.~\cite{Kabashima2026JSTAT}. For a general spectrum, $v_k$ differs from $a_k$, because the nonlinear function $\RR$ mixes band $k$ with the response $\chi_{k+1}$ of the faster bands,
and the time dependence of $\Sigma_k$ differs from $c_k$.

\paragraph{Slow components of the local field.}
Correspondingly, the local field $h$ (for SyncMC, the part of $x$ without the instantaneous self-coupling and $\eta$, which belong to the fastest band) is the sum
\begin{equation}
h(t)=H+z_{-1}+\sum_{k=0}^{K}u_k(t),\qquad u_k(t)=\phi_k(t)+\int^t\Gamma_k(t-s)\,\bar\sigma(s)\,ds ,
\label{eq:uk}
\end{equation}
where $H$ is the external field and $\bar\sigma(s)$ is the spin averaged over the variables faster than band $k$.
Let $y_k=H+z_{-1}+\sum_{j<k}u_j$ be the sum of the components slower than $u_k$. By assumption (i), $y_k$ is frozen while $u_k$ evolves,
and the faster variables follow the current value of $u_k$. We define $\Phi_{k+1}(y)$ as the free energy obtained by integrating out $u_{k+1},\dots,u_K$ and the spin at a fixed slow field $y$,
normalized so that the mean of $\sigma$ at fixed $y$ is $\partial_y\Phi_{k+1}(y)$.
For the fastest level, $\Phi_{K+1}(y)=\beta^{-1}\ln2\cosh\beta y$ for both dynamics (Sec.~\ref{sec:slow}). Then
\begin{equation}
\bar\sigma(s)=F_k\big(u_k(s)\big),\qquad F_k(u)=\partial_u\Phi_{k+1}(y_k+u),
\label{eq:sigmabar}
\end{equation}
and Eq.~\eqref{eq:uk} becomes a closed equation for $u_k$, which depends on its whole past through $\Gamma_k$.

\paragraph{Markovian form.}
Suppose first that the noise of band $k$ decays exponentially, $\Sigma_k(\tau)=v_k\,e^{-|\tau|/\tau_k}$. By Eq.~\eqref{eq:Sigmak}, $\Gamma_k(\tau)=(w_k/\tau_k)\,e^{-\tau/\tau_k}$ for $\tau>0$.
The retarded term $m_k(t)=\int^t\Gamma_k(t-s)\bar\sigma(s)\,ds$ then obeys $\tau_k\dot m_k=-m_k+w_k\bar\sigma$,
and $\phi_k$ can be generated as the Ornstein--Uhlenbeck process $\tau_k\dot\phi_k=-\phi_k+\sqrt{2v_k\tau_k}\,\xi$, where $\xi$ is white Gaussian noise with $\langle\xi(t)\xi(s)\rangle=\delta(t-s)$.
Adding the two and using Eq.~\eqref{eq:sigmabar}, we obtain the overdamped Langevin equation
\begin{equation}
\tau_k\,\dot u_k=-w_k\,\partial_uU_k(u_k)+\sqrt{2v_k\tau_k}\;\xi(t),\qquad
U_k(u)=\frac{u^2}{2w_k}-\Phi_{k+1}(y_k+u).
\label{eq:Langevin_k}
\end{equation}
In discrete time (SyncMC steps, or MCS for asynchronous MH), the same steps give the update $u_k(t+1)=\lambda u_k(t)+w_k(1-\lambda)F_k(u_k(t))+\sqrt{v_k(1-\lambda^2)}\,\zeta(t)$
with $\lambda=e^{-1/\tau_k}$ and independent standard Gaussian variables $\zeta(t)$, which reduces to Eq.~\eqref{eq:Langevin_k} for $\tau_k\gg1$.
This form exists because, by assumption (ii) and Eq.~\eqref{eq:Sigmak}, the noise and the retarded self-interaction of band $k$ have the same time dependence.

\paragraph{Stationary distribution and recursion.}
In Eq.~\eqref{eq:Langevin_k} the noise is additive and the force is the gradient of $U_k$, so the stationary distribution satisfies detailed balance and has the temperature
$v_k/w_k=T/x_k$:
\begin{equation}
P_k(u)\propto\exp\big[-\beta x_kU_k(u)\big]=\exp\Big(-\frac{u^2}{2v_k}\Big)\,Z_{k+1}(y_k+u)^{x_k},\qquad Z_{k+1}=e^{\beta\Phi_{k+1}},
\label{eq:Pk}
\end{equation}
where we used $\beta x_k/w_k=1/v_k$ from Eq.~\eqref{eq:wkvk}. The normalization gives the free energy felt by the next slower band,
\begin{equation}
Z_{K+1}(y)=2\cosh\beta y,\qquad
Z_k(y)=\Big\{\E_{z}\big[Z_{k+1}(y+\sqrt{v_k}\,z)^{x_k}\big]\Big\}^{1/x_k},
\label{eq:recursion_general}
\end{equation}
where $\E_z$ is the average over $z\sim\mathcal N(0,1)$. This has the form of Parisi's recursion, with the variance $a_k$ of the SK model replaced by $v_k$ of Eq.~\eqref{eq:wkvk}.
The plateau values follow from $q_k=\E\big[(\partial_y\Phi_k(y_k))^2\big]$, where the average is over $y_k$ distributed according to the static field $z_{-1}$ and the reweighted distributions $P_j$ of the slower bands $j<k$.
Together with the marginal-stability conditions that fix the $x_k$, these equations determine the slow-time-scale solution.
For the ROM with one slow band ($K=1$, $q_0=0$, $x_0=x$, $q_1=q$), Eq.~\eqref{eq:wkvk} gives $v_0=[\RR(\beta(1-q)+\beta xq)-\RR(\beta(1-q))]/(\beta x)=\Sigma_q$,
and Eq.~\eqref{eq:Pk} is the reweighting by $(2\cosh\beta h)^x$ used in Sec.~\ref{sec:slow}.

\paragraph{Other time dependences.}
If $\Sigma_k(\tau)$ is a superposition of decaying exponentials with positive weights, $\Sigma_k(\tau)=\sum_j\rho_j\,v_k\,e^{-|\tau|/\tau_{k,j}}$ with $\sum_j\rho_j=1$,
one introduces a variable $u_{k,j}$ for each term, with variance $\rho_jv_k$ and static response $\rho_jw_k$.
The variables then perform a multivariable overdamped Langevin motion with energy $\sum_ju_{k,j}^2/(2\rho_jw_k)-\Phi_{k+1}(y_k+\sum_ju_{k,j})$,
and the ratio of noise to friction is $T/x_k$ for every component because of Eq.~\eqref{eq:Sigmak}.
The marginal distribution of $u_k=\sum_ju_{k,j}$ is again Eq.~\eqref{eq:Pk}.
More generally, any noise whose spectrum $\tilde\Sigma_k(\omega)$ is nonnegative, as required for a covariance, can be generated by a set of harmonic oscillators coupled linearly to $u_k$~\cite{Zwanzig1973};
the oscillators then have kinetic energy, which does not change the distribution of $u_k$ at temperature $T/x_k$.
For the SK model with exponential $c_k$, $\Sigma_k$ is exponential. For other spectra, $\Sigma_k$ is not exponential even when $c_k$ is, but the argument above does not depend on its shape.

\paragraph{Role of assumption (ii).}
For any memory kernel and noise covariance that are sums of exponentials, a Markovian rewriting with auxiliary variables is possible.
Assumption (ii) is what makes the rewritten system move in the potential $U_k$ at the single temperature $T/x_k$, so that it satisfies detailed balance and has the stationary distribution of Eq.~\eqref{eq:Pk}.
The derivation shows that this distribution is consistent with (ii); it does not derive (ii) from the effective process, and it does not determine $x_k$.

\section{Alternative method: continuous-time sampling of the effective process of asynchronous MH}
\label{app:kink}

This appendix describes a way to solve the effective process of asynchronous MH that avoids the discretization of the flip times in the Bernoulli scheme of Sec.~\ref{sec:numerics}.
It was not used for the results of this paper.

\paragraph{Representation of trajectories.}
A trajectory $\sigma(t)$ of the effective process of asynchronous MH is constant between flips.
It is therefore specified exactly by its initial value $\sigma(0)$ and the list of flip times $\tau_1<\tau_2<\cdots$.
Because $W\le1$, the number of flips per unit time is at most one on average, so a trajectory up to time $L$ is stored with $O(L)$ numbers, independently of any time step.

\paragraph{Exact sampling by thinning.}
Because the flip rate is bounded by one, the trajectory can be sampled without discretizing time by the thinning method for point processes with time-dependent rates~\cite{LewisShedler1979,Ogata1981}:
\begin{enumerate}
\item Generate candidate times $\tau^{\rm c}_1<\tau^{\rm c}_2<\cdots$ from a Poisson process with rate one.
\item At each candidate time, evaluate the field $h(\tau^{\rm c})$ and flip the spin with probability $W(\sigma(\tau^{\rm c-}),h(\tau^{\rm c}))$, where $\tau^{\rm c-}$ denotes the instant just before $\tau^{\rm c}$.
\end{enumerate}
The resulting trajectory is distributed exactly according to the continuous-time jump process with rate $W(\sigma(t),h(t))$, given the field $h(t)$.
The candidate times do not depend on the state of the spin and can be generated in advance.
The field is evaluated only at the candidate times, about once per unit time, instead of $1/\Delta t$ times per unit time.
This also gives a simple picture of the effective process: a single spin that receives an MH update at random instants, once per MCS on average.
In SyncMC, by contrast, the spin is updated at every integer time.

\paragraph{Retarded self-interaction.}
Since $\sigma$ is constant between flips, the retarded term reduces to a sum over the intervals between flips.
With the cumulative kernel $K_\Gamma(t,s)=\int_0^s\Gamma_{\rm MH}(t,u)\,du$, stored on a grid,
\begin{equation}
\int_0^t\Gamma_{\rm MH}(t,s)\,\sigma(s)\,ds=\sum_{m}\sigma_m\big[K_\Gamma(t,s_{m+1})-K_\Gamma(t,s_m)\big],
\end{equation}
where $s_0=0$, $s_m=\tau_m$ for $m\ge1$ up to the last flip before $t$, the last upper limit is $t$, and $\sigma_m$ is the value of the spin on the interval $[s_m,s_{m+1})$.
The cost of one evaluation is proportional to the number of flips before $t$ and does not depend on any time step.

\paragraph{Correlation and response functions.}
The two-time functions $C$, $G$, $\Gamma_{\rm MH}$ and $\Sigma_\phi$ are stored on a grid with spacing $\Delta t_K$.
Apart from the cusp of $C$ and the jump of $G$ at equal times, which can be treated separately, these functions vary on time scales of order one,
so $\Delta t_K$ can be chosen independently of, and coarser than, the step $\Delta t$ needed by the Bernoulli discretization.
The correlation function on the grid is computed exactly from the lists of flip times.
For the response function, we use the derivative of the log path probability with respect to a field applied during the grid interval $l$.
Given the field, the probability density of a trajectory with flips at $\tau_1<\tau_2<\cdots$ is $\prod_mW(\sigma(\tau_m^-),h(\tau_m))\exp[-\int W(\sigma(t),h(t))\,dt]$ (Appendix~\ref{app:cont}), so
\begin{equation}
S_l=\sum_{\substack{\text{flips}\\\text{in interval }l}}\partial_h\log W\big(\sigma(\tau_m^-),h(\tau_m)\big)-\int_{\text{interval }l}\partial_hW\big(\sigma(t),h(t)\big)\,dt,
\label{eq:score_kink}
\end{equation}
and $G(t_k,t_l)=\langle\sigma(t_k)\,S_l\rangle/\Delta t_K$ for $k>l$.
For the Metropolis rate, $\partial_h\log W=-2\beta\sigma$ and $\partial_hW=-2\beta\sigma W$ if $\sigma h>0$, and both vanish otherwise.
The first term is computed exactly from the list of flips.
For the integral, $\sigma(t)$ is known at all times, and $h(t)$ is known at the grid points and at the flip times, where the noise has been sampled.
We split the interval at the flip times and apply the trapezoidal rule on each piece, with $\sigma$ set to its value on that piece.
The error is of order $\Delta t_K^2$ per interval and vanishes as $\Delta t_K\to0$, like the error of the noise.
An unbiased estimate is also possible: generate auxiliary times from a Poisson process with rate $\lambda$, independently of the trajectory, sample the noise there in the same way, and replace the integral by $\lambda^{-1}\sum\partial_hW$ over these times.
The candidate times cannot be used for this purpose, because the flips are among them.
Given the flip times, the rejected candidates form a Poisson process with rate $1-W$, so the sum of $\partial_hW$ over all candidates estimates $\sum_m\partial_hW(\tau_m)+\int(1-W)\,\partial_hW\,dt$ rather than $\int\partial_hW\,dt$, and the resulting estimate of $G$ is biased.
An alternative estimator, obtained by differentiating the probability of each accept or reject decision, contains the term $\partial_hW/(1-W)$ for rejected candidates;
since $1-W\simeq2\beta|h|$ as $\sigma h\to0^+$, its variance diverges logarithmically, and it should not be used with the Metropolis rate.

\paragraph{Gaussian noise at the candidate times.}
The noise $\phi$ is needed at the candidate times, which differ from one trajectory to another.
Sharing the candidate times among trajectories would allow a single factorization of the covariance matrix, but it is not correct:
in the true dynamics the candidate times of different sites are independent, and sharing them would make the measured correlation function conditional on a particular set of candidate times
(for example, $C$ would be exactly one between consecutive candidate times).
We therefore generate $\phi$ on the grid of spacing $\Delta t_K$, using one factorization of $\Sigma_\phi$ for all trajectories,
and obtain the values at the candidate times by sampling from the Gaussian distribution conditioned on the values at the neighboring grid points.
Because $\Sigma_\phi$, like $C$, has a cusp at equal times, $\phi$ behaves locally like an Ornstein--Uhlenbeck process, for which this conditional sampling is accurate.
The remaining error vanishes as $\Delta t_K\to0$ and can be removed by extrapolation.

\paragraph{Comparison with continuous-time quantum Monte Carlo.}
The representation by flip times is analogous to the kink (segment) representation of continuous-time quantum Monte Carlo methods.
The difference is that here the trajectories are sampled directly in real time by thinning,
whereas in quantum Monte Carlo configurations of kinks in imaginary time are sampled by a Markov chain according to their weights.

Table~\ref{tab:kink} compares the costs of the two approaches.

\begin{table}[t]
\caption{Costs of solving the effective process of asynchronous MH up to time $L$ by the Bernoulli discretization with step $\Delta t$ and by the flip-time representation with thinning. $\Delta t_K$ is the grid spacing for the two-time functions; in the flip-time representation the noise at the candidate times is obtained by conditional sampling from its values on this grid.}
\label{tab:kink}
\begin{ruledtabular}
\begin{tabular}{p{0.36\linewidth}p{0.24\linewidth}p{0.3\linewidth}}
 & Bernoulli discretization & Flip times and thinning\\
\hline
Field evaluations per trajectory & $L/\Delta t$ & about $L$\\
Cost per evaluation of retarded term & $O(L/\Delta t)$ & $O(\text{number of flips})$\\
Error in the flip times & $O(\Delta t)$ & none\\
Size of the two-time matrices & $(L/\Delta t)^2$ & $(L/\Delta t_K)^2$\\
Error in the noise & none on the grid & vanishes as $\Delta t_K\to0$\\
\end{tabular}
\end{ruledtabular}
\end{table}

\end{document}